\documentclass[twocolumn]{aastex61}
\usepackage[english]{babel}
\usepackage{gensymb}
\usepackage{textcomp}
\graphicspath{{images/}}

\received{December 14, 2017}
\submitjournal{Publications of the Astronomical Society of the Pacific}

\shorttitle{ProtoDESI: First On-Sky Demonstration for DESI}
\shortauthors{Fagrelius et al.}

\begin{document} 

    \title{ProtoDESI: First On-Sky Technology Demonstration for the Dark Energy Spectroscopic Instrument}


    \author{Parker Fagrelius}
    \affil{Department of Physics, University of California, Berkeley, Berkeley, CA 94720, USA}

    \author{Behzad Abareshi}
    \affil{National Optical Astronomy Observatory, Tucson, AZ 85719, USA}

    \author{Lori Allen}
    \affil{National Optical Astronomy Observatory, Tucson, AZ 85719, USA}

    \author{Otger Ballester}
    \affil{Institut de F\'{i}sica d'Altes Energies (IFAE), The Barcelona Institute of Science and Technology, Campus UAB, 08193 Bellaterra (Barcelona) Spain}

    \author{Charles Baltay}
    \affil{Astronomy Department, Yale University, New Haven, CT 06520, USA}

    \author{Robert Besuner}
    \affil{Department of Physics, University of California, Berkeley, Berkeley, CA 94720, USA}

    \author{Elizabeth Buckley-Geer}
    \affil{Fermi National Accelerator Laboratory, Batavia, IL 60510, USA}

    \author{Karen Butler}
    \affil{National Optical Astronomy Observatory, Tucson, AZ 85719, USA}

    \author{Laia Cardiel}
    \affil{Institut de F\'{i}sica d'Altes Energies (IFAE), The Barcelona Institute of Science and Technology, Campus UAB, 08193 Bellaterra (Barcelona) Spain}

    \author{Arjun Dey}
    \affil{National Optical Astronomy Observatory, Tucson, AZ 85719, USA}
    
    \author{Yutong Duan}
    \affil{Physics Department, Boston University, Boston, MA 02215, USA}

    \author{Ann Elliott}
    \affil{Department of Astronomy,  The Ohio State University, Columbus, OH 43210, USA}

    \author{William Emmet}
    \affil{Astronomy Department, Yale University, New Haven, CT 06520, USA}

    \author{Irena Gershkovich}
    \affil{Department of Astronomy, University of Michigan, Ann Arbor, MI 48109, USA}

    \author{Klaus Honscheid}
    \affil{Department of Astronomy,  The Ohio State University, Columbus, OH 43210, USA}

    \author{Jose M. Illa}
    \affil{Institut de F\'{i}sica d'Altes Energies (IFAE), The Barcelona Institute of Science and Technology, Campus UAB, 08193 Bellaterra (Barcelona) Spain}

    \author{Jorge Jimenez}
    \affil{Institut de F\'{i}sica d'Altes Energies (IFAE), The Barcelona Institute of Science and Technology, Campus UAB, 08193 Bellaterra (Barcelona) Spain}

    \author{Richard Joyce}
    \affil{National Optical Astronomy Observatory, Tucson, AZ 85719, USA}

    \author{Armin Karcher}
    \affil{Lawrence Berkeley National Laboratory, Berkeley, CA 94720, USA}

    \author{Stephen Kent}
    \affil{Fermi National Accelerator Laboratory, Batavia, IL 60510, USA}

    \author{Andrew Lambert}
    \affil{Lawrence Berkeley National Laboratory, Berkeley, CA 94720, USA}

    \author{Michael Lampton}
    \affil{Lawrence Berkeley National Laboratory, Berkeley, CA 94720, USA}

    \author{Michael Levi}
    \affil{Lawrence Berkeley National Laboratory, Berkeley, CA 94720, USA}

    \author{Christopher Manser}
    \affil{Department of Physics, University of Warwick, Coventry, UK}

    \author{Robert Marshall}
    \affil{National Optical Astronomy Observatory, Tucson, AZ 85719, USA}

    \author{Paul Martini}
    \affil{Department of Astronomy,  The Ohio State University, Columbus, OH 43210, USA}

    \author{Anthony Paat}
    \affil{National Optical Astronomy Observatory, Tucson, AZ 85719, USA}

    \author{Ronald Probst}
    \affil{National Optical Astronomy Observatory, Tucson, AZ 85719, USA}

    \author{David Rabinowitz}
    \affil{Astronomy Department, Yale University, New Haven, CT 06520, USA}

    \author{Kevin Reil}
    \affil{SLAC National Accelerator Laboratory, Menlo Park, CA 94025, USA}

    \author{Amy Robertson}
    \affil{National Optical Astronomy Observatory, Tucson, AZ 85719, USA}

    \author{Connie Rockosi}
    \affil{University of California Observatories, Santa Cruz, CA 95065, USA}

    \author{David Schlegel}
    \affil{Lawrence Berkeley National Laboratory, Berkeley, CA 94720, USA}

    \author{Michael Schubnell}
    \affil{Department of Astronomy, University of Michigan, Ann Arbor, MI 48109, USA}

    \author{Santiago Serrano}
    \affil{Institut de Ci\`encies de l'Espai, IEEC-CSIC, Campus UAB, Carrer de Can Magrans s/n, Barcelona, Spain}

    \author{Joseph Silber}
    \affil{Lawrence Berkeley National Laboratory, Berkeley, CA 94720, USA}

    \author{Christian Soto}
    \affil{National Optical Astronomy Observatory, Tucson, AZ 85719, USA}

    \author{David Sprayberry}
    \affil{National Optical Astronomy Observatory, Tucson, AZ 85719, USA}

    \author{David Summers}
    \affil{National Optical Astronomy Observatory, Tucson, AZ 85719, USA}

    \author{Greg Tarl\'{e}}
    \affil{Department of Astronomy, University of Michigan, Ann Arbor, MI 48109, USA}
    
    \author{Benjamin A. Weaver}
    \affil{National Optical Astronomy Observatory, Tucson, AZ 85719, USA}

    \begin{abstract}
        {The Dark Energy Spectroscopic Instrument (DESI) is under construction to measure the expansion history of the universe using the baryon acoustic oscillations technique. The spectra of 35 million galaxies and quasars over 14,000 square degrees will be measured during a 5-year survey. A new prime focus corrector for the Mayall telescope at Kitt Peak National Observatory will deliver light to 5,000 individually targeted fiber-fed robotic positioners. The fibers in turn feed ten broadband multi-object spectrographs. We describe the ProtoDESI experiment, that was installed and commissioned on the 4-m Mayall telescope from August 14 to September 30, 2016. ProtoDESI was an on-sky technology demonstration with the goal to reduce technical risks associated with aligning optical fibers with targets using robotic fiber positioners and maintaining the stability required to operate DESI. The ProtoDESI prime focus instrument,  consisting of three fiber positioners, illuminated fiducials, and a guide camera, was installed behind the existing Mosaic corrector on the Mayall telescope. A Fiber View Camera was mounted in the Cassegrain cage of the telescope and provided feedback metrology for positioning the fibers. ProtoDESI also provided a platform for early integration of hardware with the DESI Instrument Control System that controls the subsystems, provides communication with the Telescope Control System, and collects instrument telemetry data. Lacking a spectrograph, ProtoDESI monitored the output of the fibers using a Fiber Photometry Camera mounted on the prime focus instrument. ProtoDESI was successful in acquiring targets with the robotically positioned fibers and demonstrated that the DESI guiding requirements can be met.}
    \end{abstract}

    \keywords{dark energy: BAO -- instrumentation: miscellaneous -- methods: observational -- telescopes  }

    \section{Introduction}
        \label{sec:intro}
        
        The Dark Energy Spectroscopic Instrument (DESI) is a dark energy experiment that will create a 3D map of the universe to a redshift of $z\sim3.5$, measuring the spectra from 35 million galaxies and quasars. This map of the large scale structure of the universe will enable the measurement of Baryon Acoustic Oscillations (BAO), a standard ruler that emerges from primordial fluctuations in the early universe \citep{seo, weinberg}. Using BAO and Redshift Space Distortions (RSD) \citep{kaiser}, the DESI project will constrain knowledge of the time evolution of dark energy, responsible for the acceleration of cosmic expansion and making up $\sim70\%$ of the energy density in the current universe. DESI is expected to improve the Dark Energy Figure of Merit (DETF; \citealt{DETF}) by up to 10 times that of the last generation of dark energy experiments \citep{SDR}, such as the Baryon Acoustic Oscillation Spectroscopic Survey (BOSS; \citealt{BOSS}). DESI, a Stage IV Dark Energy experiment in the parlance of the DETF report, will revolutionize our understanding of the universe.
    
        To achieve these science goals, DESI employs 5,000 robotic fiber positioners to simultaneously place optical fibers on astronomical targets. The focal plate instrument, containing the robotic positioners, will sit behind a custom optical corrector, both of which will be installed at the primary focus of the 4-m primary mirror of the NOAO (National Optical Astronomy Observatory) Mayall telescope. The optical fibers will run from the robotic positioners at the prime focus down the telescope to ten 3-arm spectrographs, which cover a spectral range of 360-980 nm with a resolution of 2,000-5,000 \citep{IDR}. This enables DESI to probe redshifts up to 1.7 for emission line galaxies and to 3.5 for Lyman-$\alpha$ spectra from quasars. With exposure times of $\approx$20 minutes, DESI will be capable of measuring the spectra of up to 100,000 galaxies in a night with a signal-to-noise ratio (SNR) of $\approx$10 for Emission Line Galaxy [OII] doublets.  DESI will be installed at the Mayall telescope at Kitt Peak National Observatory (KPNO) in 2018-2019, followed by a dedicated 5 year science campaign.
    
        In order to measure the spectra with the 107 $\mu$m diameter DESI optical fibers, sampling the focal plane with an average plate scale of $\approx$0.014 arcseconds/$\mu$m, the robotic positioners must arrive at their target with $\leq$10 $\mu$m RMS error and maintain pointing for the duration of the DESI observation. This level of precision had been tested in the lab, but not fully demonstrated on-sky with positioners and other key DESI components integrated together. In order to reduce the risk related to these technical challenges, a prototype instrument was built and commissioned on the Mayall 4-m telescope in the summer of 2016. This prototype, called ProtoDESI, is a subset of DESI subsystems that was used for the first on-sky test of critical functions for the larger DESI instrument. 
        
        ProtoDESI included three robotic positioners, a guiding camera, a fiber metrology camera, and a subset of the instrument control software. It did not include long optical cables or spectrographs, like the final DESI Instrument will, but it did include shorter fibers that fed an imager for monitoring alignment with targets. The focal plate instrument was small enough to fit in the existing prime focus bay of the Mayall telescope with the existing corrector. During the time installed on the KPNO Mayall telescope from August 14, 2016 to September 30, 2016, ProtoDESI demonstrated that the Mayall telescope could be guided using the DESI architecture and that the DESI robotic positioners moved such that light from targets could be captured by the DESI optical fibers and remain stable. Additionally, the commissioning of ProtoDESI required early and careful integration of many critical subsystems in the DESI design, aiding in the completion of DESI and informing commissioning plans. 
        
        On-sky subsystem testing for large complex, instruments is a common and useful practice as demonstrated by projects such as: 2MASS \citep{2mass}, HETDEX \citep{hetdex}, and DECam for the Dark Energy Survey \citep{precam}. This approach will become even more critical as the size and complexity of cosmology instruments increases, and has been adopted by LSST \citep{lsst_comcam}. Most of the commissioning instruments referenced above were used for early testing of almost-complete facilities. ProtoDESI is unique in that it represents an early on-sky proof of concept for operations and a testbed for new technologies prior to a telescope upgrade, an approach that aims to reduce risks during the development stage.

         This paper describes in detail the design of the ProtoDESI instrument as-installed (Sec. \ref{sec:instr}). It then outlines the subsystem tests completed in the lab and on the telescope, which provides a benchmark for performance (Sec. \ref{sec:commish}). When all subsystems demonstrated compliance with their design requirements, we completed an operations campaign that tested ProtoDESI's ability to guide the telescope and align fibers with targets. We describe the operations in Sec. \ref{ops} and the results in Sec. \ref{sec:results}. Finally, we discuss key lessons learned from ProtoDESI that will be considered in the DESI integration and commissioning phases (Sec. \ref{sec:disc}).  
    
        \begin{figure*}
            \centering
            \includegraphics[width=15cm]{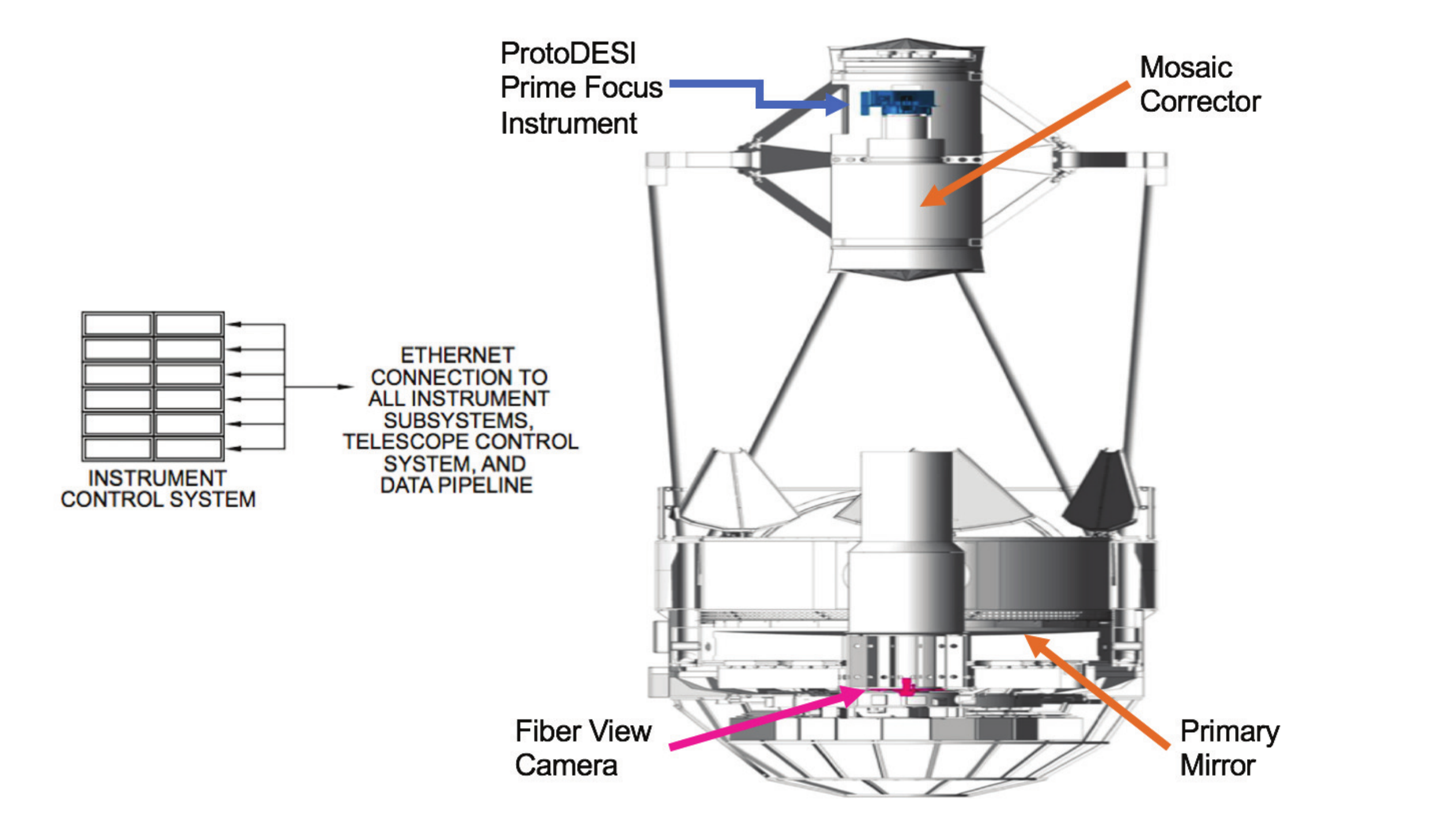}
            \caption{Layout of the ProtoDESI experiment. The prime focus instrument (blue) is mounted behind the Mosaic corrector and the FVC (pink) is mounted behind the primary mirror in the Cassegrain cage. ProtoDESI also includes the ICS and TCS.}
            \label{install1}
        \end{figure*}
    
    \section{Instrument Description}
        
        \label{sec:instr}
\begin{figure*}
            \label{primefocusinst}
            \centering
            \includegraphics[width=12cm]{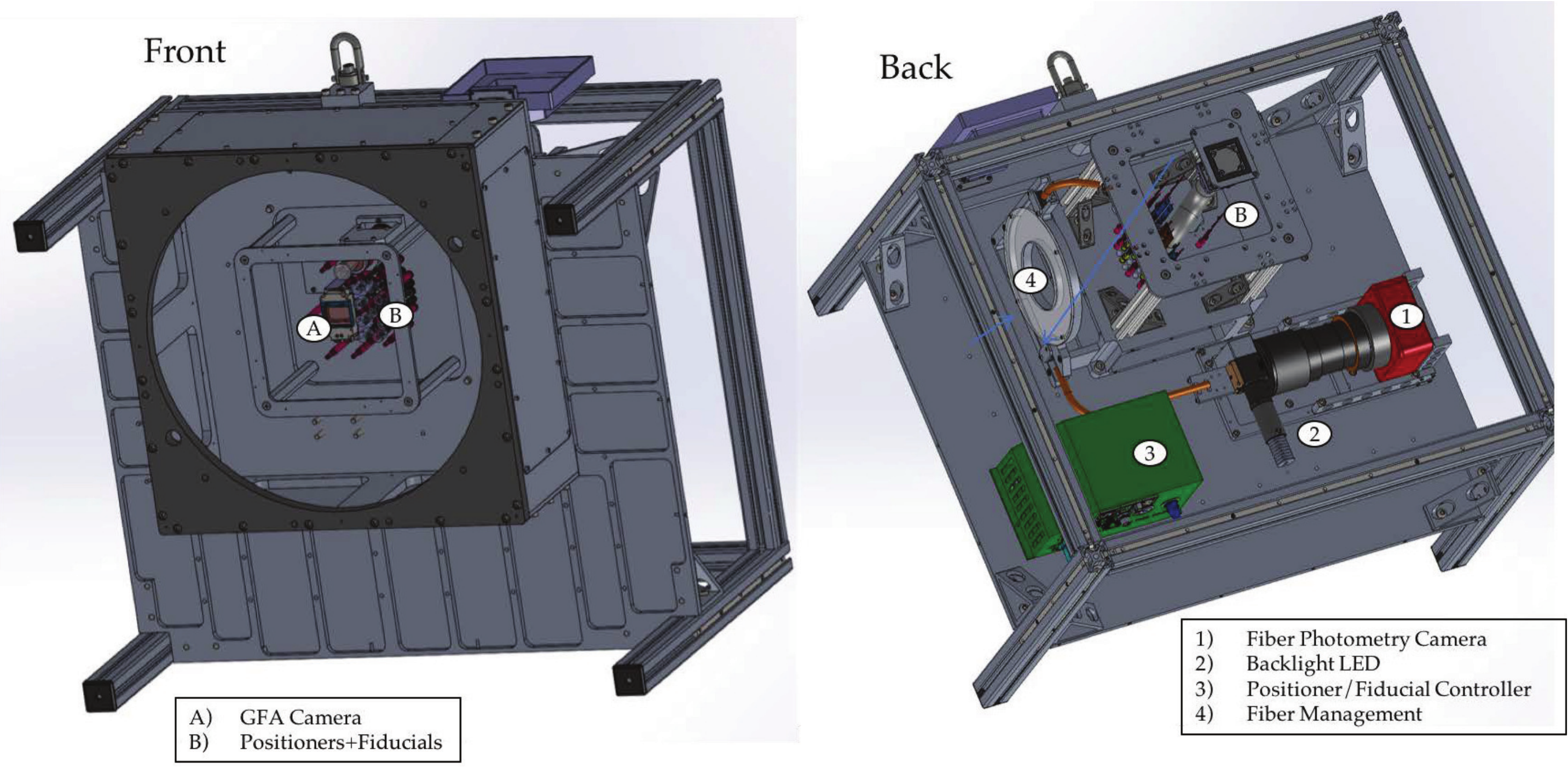}
            \caption{Computer-aided design (CAD) model of the ProtoDESI prime focus instrument. Light was incident on the front (left image) of the focal plate, which was mounted directly to the Mosaic corrector.}
            \label{focalplate}
        \end{figure*}

        The ProtoDESI instrument is a subset of DESI subsystems, including a much simplified prime focus instrument with three positioners, fiducials and a Guide, Focus and Alignment (GFA) camera, a Fiber View Camera (FVC) for positioning feedback, the Telescope Control System (TCS), and the DESI Instrument Control System (ICS) (Fig. \ref{install1}). Rather than measuring the output of a 50-meter fiber cable with the DESI spectrographs, ProtoDESI confirmed fiber pointing and stability with a fiber photometry camera (FPC) that imaged the ends of the short 3-meter fibers mounted on the prime focus instrument. The prime focus instrument (Fig. \ref{focalplate}), which included the fibers and the robotic fiber positioners, illuminated fiducials, GFA and the FPC, sat behind the existing Mosaic corrector \citep{doi:10.1117/12.316845}. The FVC was installed in the Cassegrain cage of the Mayall telescope where it could image the front of the focal plane. The following gives a detailed description of the instrument as-installed.
        
        \subsection{Prime Focus Instrument}
        
            \label{sec:pf}
            The prime focus instrument was designed so that it could be installed like the Mosaic prime-focus camera, requiring no modifications to the telescope. The instrument is $750 \times 800 \times 500$ mm with a total mass of $\approx$80 kg, and was mounted behind the f/3.2 refractive Mosaic corrector which has a 0.8 degree field of view (FOV). The flat focal plate held 16 fiducials (point sources fixed to the focal plate), three robotic positioners, one GFA camera, and a backup finder scope camera. The focal plate was mounted to a support structure which housed the fiber management system, FPC, and all power and control electronics. The instrument received all power and network connections in the prime focus bay via a single feedthrough in the instrument. The front mounting plate of the ProtoDESI instrument (Fig. \ref{focalplate}, left) was machined to match the mounting surface of the Mosaic3 wide-field imager \citep{Mosaic3}. Therefore, it could be installed directly to the corrector using the same lifting and installation procedures used for Mosaic3, using pins to accurately align the focal plate before securing it.
            
            \subsubsection{Robotic Positioners}
\begin{figure*}
                \centering
                \includegraphics[width=15cm]{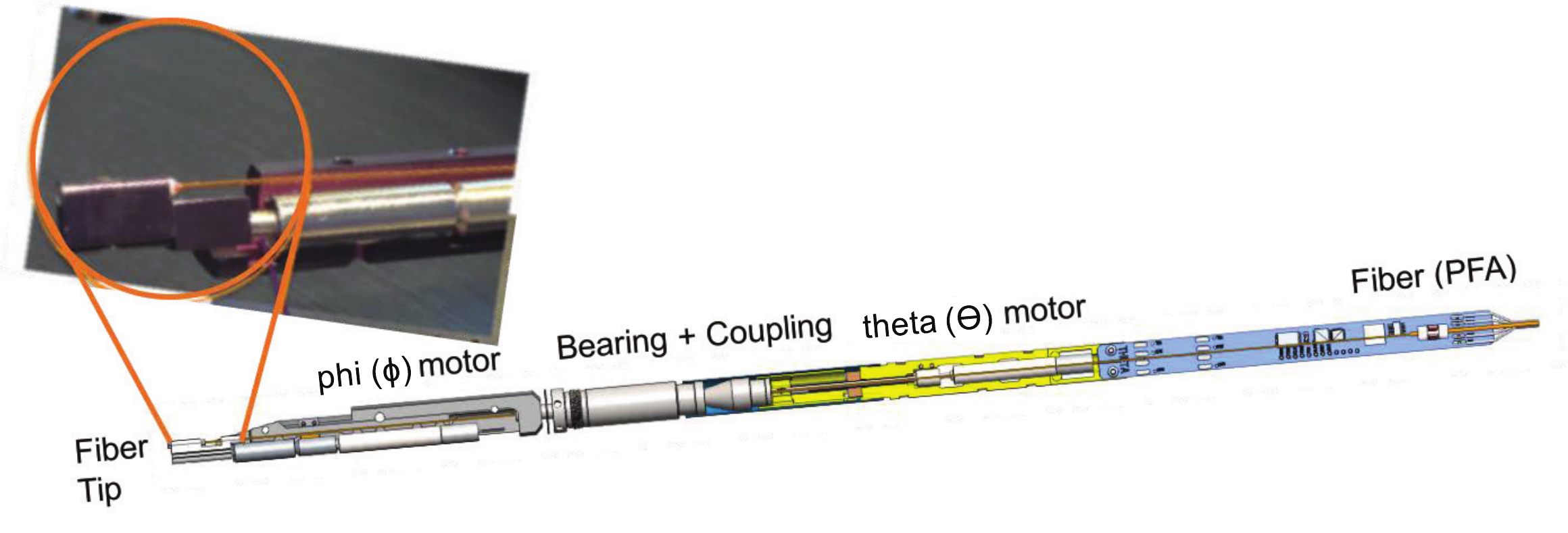}
                \caption{The DESI robotic positioner. Magnified is the fiber tip, glued into a glass ferrule and secured to the positioner with a set screw. The positioner measures 29 cm in length from the ferrule to the bottom of the custom electronics board.}
                \label{positioner}
            \end{figure*}

            The DESI robotic positioners are used to reconfigure the focal plane, gaining us access to new targets on a short time scale. Each positioner holds a single optical fiber, and has two 4 mm DC brushless motors, enabling it to access any point within a 6 mm radius patrol disk. The motors are arranged so that there are two rotational degrees of freedom with parallel axes \citep{spie_positioners}. This style of positioner is often referred to as theta-phi, where the central motor defines the $\theta$ axis, and the eccentric motor defines the $\phi$ axis. The motors do not include individual rotary encoders. Instead, the positioning control loop is closed by directly imaging the backlit fiber tips with the FVC (described below). The robotic positioners were built by the University of Michigan. Each unit has an external thread coaxial with a precise cylindrical datum feature, and screws into the focal plate like a sparkplug. A flange co-machined into the unit positions the fiber tip 86.5 mm forward of the plate's mechanical mounting surface. They were designed to be mounted on a grid spaced by 10.2 mm, allowing for an overfilled focal plate on DESI with patrol areas overlapping. This will require anti-collision algorithms ensuring that robots do not run into each other during reconfiguration. ProtoDESI did not include this capability so the positioners were placed such that their patrol areas did not overlap. The fiducials were mounted in the same way on the focal plate, having the exact same length so that the fibers and fiducial pinholes all lie in the same focal surface. The optical fiber was attached to the positioner with a set screw that holds a glass ferrule glued to the tip of the fiber (Fig. \ref{positioner}). Each individual positioner includes an integrated custom electronics board that powers and controls the brushless motor function.  The positioners received 7.5 V and commands were sent using the CAN bus protocol from a petal controller (PC) based on a BeagleBone Black\footnote{https://beagleboard.org/black} microcomputer. The PC also managed the thermal control system, consisting of four temperature sensors and two fans. One of the fans cooled the GFA and the other created positive pressure across the front of the focal plate to avoid deposition of dust on the optics.
        
            \subsubsection{Optical Fibers}
            Three meters of DESI fiber were attached to each positioner to carry light from the front of the focal plate to be imaged by the FPC on the back of the prime focus instrument. The optical fiber is broadband Polymicro FBP fiber with a core diameter of 107 $\mu$m and a numerical aperture of 0.22. The length of the fiber ensured that the output of the fibers was sufficiently azimuthally scrambled, avoiding any near field structure. The light cone from the Mosaic corrector to the focal plane is $\approx$f/3.2, with a platescale of 0.017 arcsec/$\mu$m, so each fiber saw 2.4 $\mathrm{arcsec}^2$ of the sky. 
            
            \subsubsection{Fiducials}
\begin{figure*}
                \centering
                \includegraphics[width=7cm]{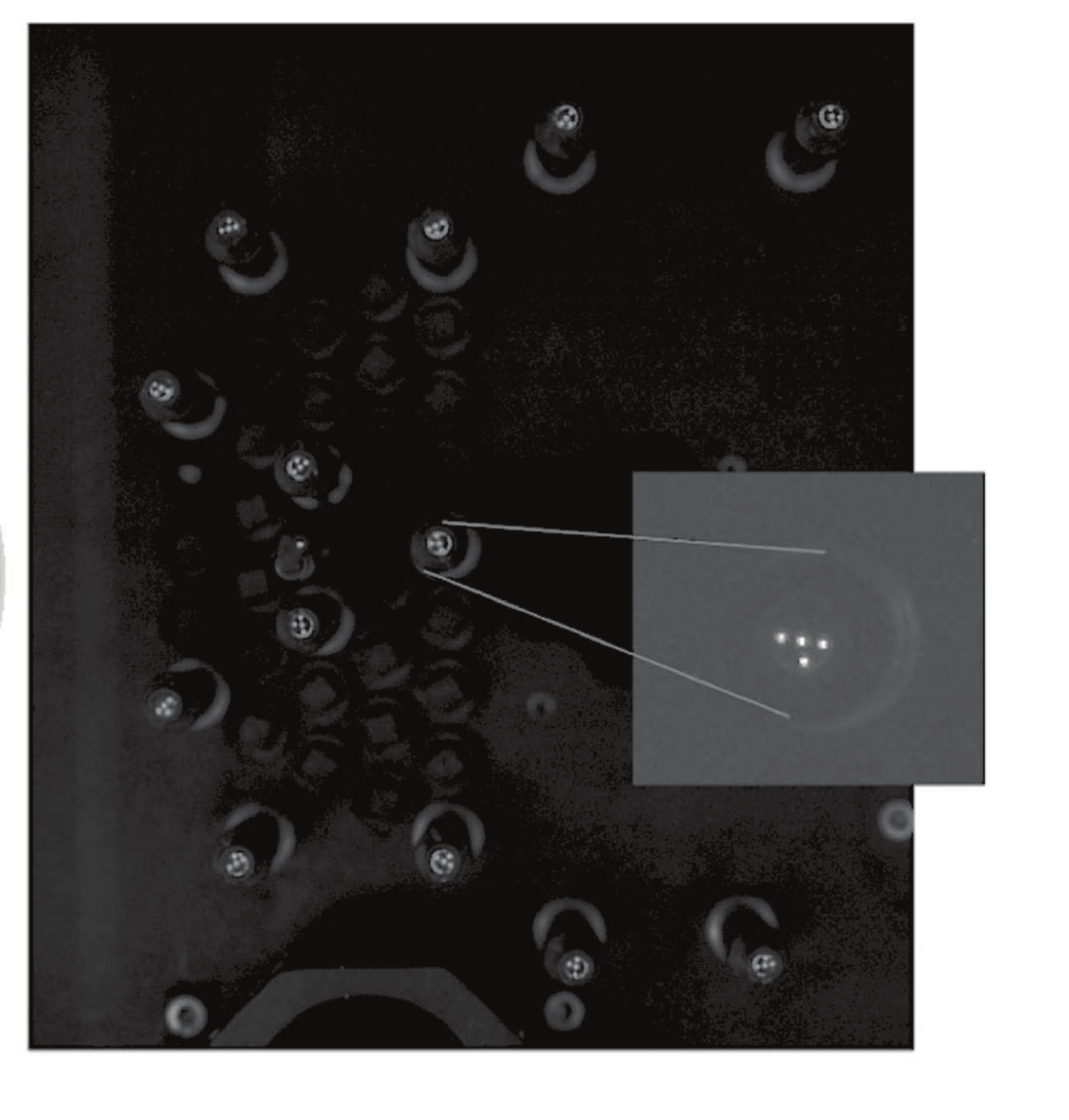}
                \includegraphics[width=8cm]{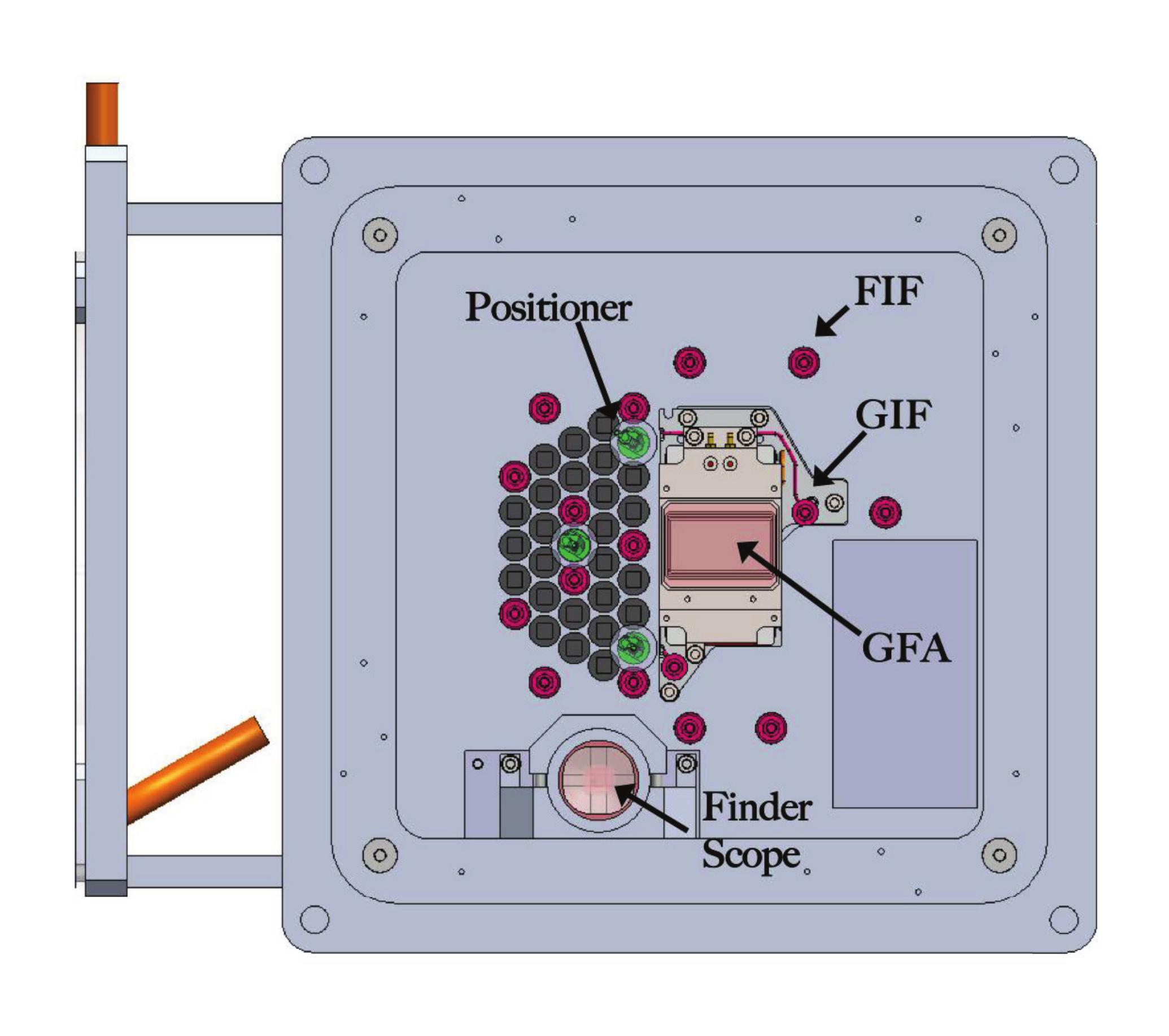}
                \caption{\textbf{Left:} Illuminated fiducials with 4 pinholes each mounted on the focal plate. \textbf{Right:} Schematic drawing of the front view of the focal plate. Locations of the fiducials (pink), fibers (green), and empty holes (black) are identified along with the GFA (rectangular camera at center) and the finder scope (circular camera towards bottom).}
                \label{fiducials}
            \end{figure*}

            The fiducials, built by Yale University, each consist of four 10 $\mu$m  pinholes etched in a black coating on a glass block that are back illuminated by a 470 nm LED (Fig. \ref{fiducials}). Their main function is to provide reference points on the focal plate relative to the movable fibers. By imaging the front of the focal plate with the FVC (See Sec.\ref{sec:FVC}) it is then possible to measure the absolute position of the science fibers through the distortions introduced by the corrector optics. The locations of the pinholes were measured relative to the outside diameter of the fiducials with an accuracy of 1.5 $\mu$m. The centers of the fiducials were measured with an accuracy of $\approx$5$\mu$m for two types of fiducials.
             Two of the 16 fiducials were mounted directly to the GFA camera body, hereafter referred to as GFA Illuminated Fiducials (GIFs). Using a metrology system in the laboratory, developed at Lawrence Berkeley National Laboratory (LBNL) for this specific purpose, the locations of the GIFs were measured relative to the GFA CCD sensor. First, the illuminated GIFs were imaged using an external camera mounted on motorized stages, and the \textit{x-y} locations of each GIF pinhole centroid was recorded. Next, an LED-generated circular dot, aligned with the center of the external camera sensor, was projected onto the GFA sensor.  The projected dot was stepped over the GFA sensor in a grid pattern, and at each step a GFA exposure was taken. The LED centroid location on the GFA sensor and the stage \textit{x-y} coordinate was recorded.  This tied the \textit{x-y} location of the GIF pinholes to the \textit{x-y} locations of GFA pixels. 
            The remaining fiducials, Field Illuminated Fiducials (FIFs), were mounted directly to the focal plate. The location of the FIFs, relative to the focal plate surface, were determined using a Coordinate Measuring Machine (CMM). After the GFA + GIF assembly was installed on the focal plate, the location of the GIFs relative to the FIFs was measured using the FVC, linking the coordinate systems of the focal plate and of the GFA sensor.  
        
        	\subsubsection{Guide, Focus and Alignment Camera}
\begin{figure}
                \centering
                \includegraphics[height=7.5cm]{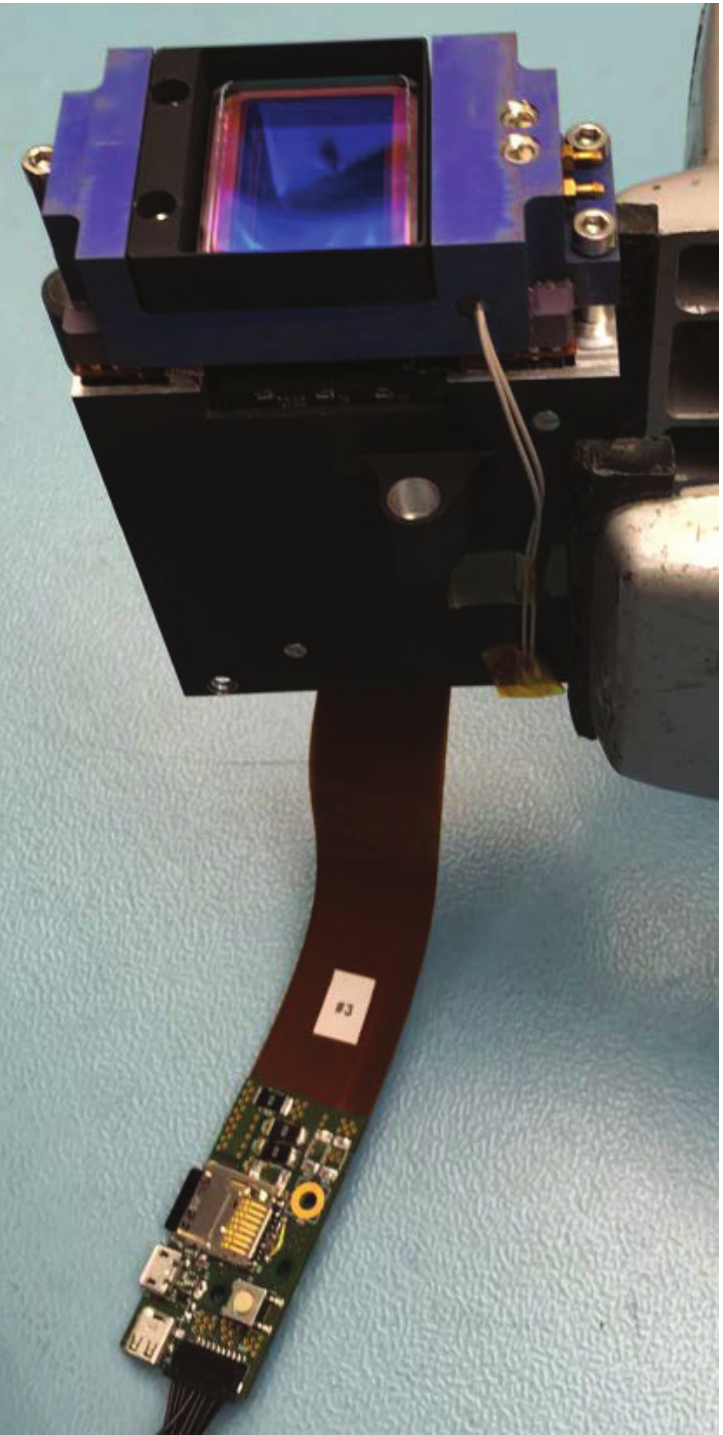}
                \caption{ProtoDESI GFA camera. The custom enclosure houses an e2v CCD with a Kapton tape cable to the electronics, giving access to power and ethernet connections.}
                \label{GFA}
            \end{figure}

            Located in the center of the focal plate, the GFA sent $\approx$30 arcmin$^{2}$ images to the ICS to be used in field identification and telescope guiding. The camera, housing an e2v CCD230-45 sensor, was custom-designed and built by LBNL and IFAE in Barcelona to minimize mechanical footprint while maximizing sensor area on the focal plate (Fig. \ref{GFA}). The GFA sensor has 15 $\mu$m pixels read out by 4 separate amplifiers. A custom SDSS r\textquotesingle-band filter was mounted to the camera to reject light from the blue fiducial LEDs and backlit fibers. The GFA required 3.3 V, 5 V, 15 V, and 32 V power inputs, supplied by linear regulated AC-DC supplies connected to an interlock so that if the FPGA on the GFA exceeded 60$^{\circ}$~C the power would be externally shut down, avoiding damage to the CCD. The cameras ran at ambient temperature and were cooled only by a fan dispersing heat behind the sensor. The data from the GFA was delivered directly over ethernet to the ICS where images were assembled for guiding. 
        
            In addition to the GFA, the focal plate contained a finder scope camera. This was to be used in the event that we were unable to align fibers with targets, we would have a way to connect images from two areas on the focal plate (finder and GFA) for troubleshooting. The finder camera selected was a SBIG\footnote{SBIG Astronomical cameras distributed by Diffraction Limited (diffractionlimited.com)}-STi Monochrome camera with a SEMI KAI-340 CCD, also with an SDSS r\textquotesingle-band filter. Besides confirming its functionality, we did not use the finder camera during operations of ProtoDESI.
        
            \subsubsection{Fiber Photometry Camera}
            
            Light from the targets ran down the fibers to the FPC where all 3 fibers were imaged in a fixed pattern. As the name suggests, the FPC provided aperture photometry on the output of the fibers, with the goal to achieve $\approx$1\% relative photometry error for each fiber. The system was designed such that the output of the fibers did not move in the FPC images, nor did they change shape (azimuthal scrambling), but as a star moved across a fiber brightness of the output would change. In addition to imaging the fibers, the FPC assembly also backlit the fibers using a beam splitter and a 470 nm LED. The LED output was adjusted using a Mightex USB/RS232 controller\footnote{Mightex Universal LED Controller (www.mightexsystems.com)} so that the fibers had the same brightness as the fiducials. The FPC is an SBIG STF-8300M containing a KAF-8300 CCD with 3326$\times$2504 pixels with a 5.4 $\mu$m pixel size. Using a Canon 50 mm f/1.4 lens, the fiber output was imaged with a magnification of 1. Like the GFA and fiber camera, the FPC also had an SDSS r\textquotesingle-band filter with 0.15 $\mu$m bandwidth to reject the 470 nm light from the fiducials and backlighting LED.  The FPC and LED were commercial-off-the-shelf parts with standard power supplies. The control software ran on an Intel NUC\footnote{Intel mini PC using Intel Core i3 Processor (https://www.intel.com/content/www/us/en/products/boards-kits/nuc.html)} that was mounted on the back of the prime focus instrument and connected to the LED and FPC via USB.
    	\newpage
        \subsection{Fiber View Camera}
        \label{sec:FVC}
            The FVC was mounted in the Cassegrain cage 12.25 m from the prime focus, oriented to image the front of the focal plane (Fig. \ref{install1}). The images of the front of the focal plate, with the fiducials illuminated and science fibers back-lit, are used for metrology. Centroids for each point imaged by the FVC provide a feedback mechanism for identifying the location of the fibers. Since the robotic positioners don't include encoders, these images are critical to ensuring that fibers arrive at their commanded locations. The FVC used for ProtoDESI was nearly identical to that which will be used for DESI \textemdash a Finger Lakes Instruments\footnote{www.flicamera.com} Proline PL501000 with a Kodak KAF50100 CCD, controlled with an Intel NUC over USB. The sensor has 6132$\times$8176 pixels with a 6 $\mu$m pitch and receives light through a blue narrow band filter. Using a Canon telephoto lens with an effective focal length of 600 mm stopped to f/19, the focal plate was de-magnified by 21.7 so that the  diffraction-limited full width half maximum (FWHM) of each fiber spanned 2 pixels, maximizing the SNR. The lens and camera were attached to an adapter plate which was installed in the Cassegrain cage, and the heavy ($\approx$4 kg) Canon lens was supported within a cage to minimize flexure due to gravitational loading.
    
        \subsection{Telescope and Telescope Control System}
        
            The Mayall telescope at KPNO is a 4-meter optical telescope on an equatorial mount. Together with its sibling, the Blanco telescope at Cerro Tololo Inter-American Observatory in La Serena, Chile, it is one of the last of its kind ever built, and will support the DESI corrector which provides a 3.2 degree FOV. It currently serves regular scientific observing, and will be converted to sole-purpose use by DESI in 2018. Before major modifications are made to accommodate DESI, ProtoDESI was designed to be mounted behind the existing Mosaic corrector.
        
            In preparation for DESI, and prior to ProtoDESI, the Mayall Telescope TCS was upgraded very similarly to the Blanco telescope in support of the Dark Energy Survey (DES; \citealt{mayall1, mayall2}). This upgrade included new encoding hardware, servos, and the telescope control software. The new system is fully digital with a programmable servo controller. Evaluation of the new system during the Mosaic z-Band Legacy Survey (MzLS) showed the RMS pointing error to be 3-4 arcseconds with a settling time after slew of less than 10 seconds, with minimal dependence on the slew angle. With these improvements, the TCS provided open loop tracking with stability better than 0.17 arcsec/min (10 $\mu$m/min on the focal plate) RMS error, exceeding expectations for ProtoDESI. Additionally, the Mayall telescope control room was moved and enlarged to meet the needs of the large DESI commissioning team, and the ProtoDESI team acted as ``beta'' testers, leading to additional improvements.
    
        \subsection{Instrument Control System}
        
\begin{figure*}
                \centering
                \includegraphics[width=12cm]{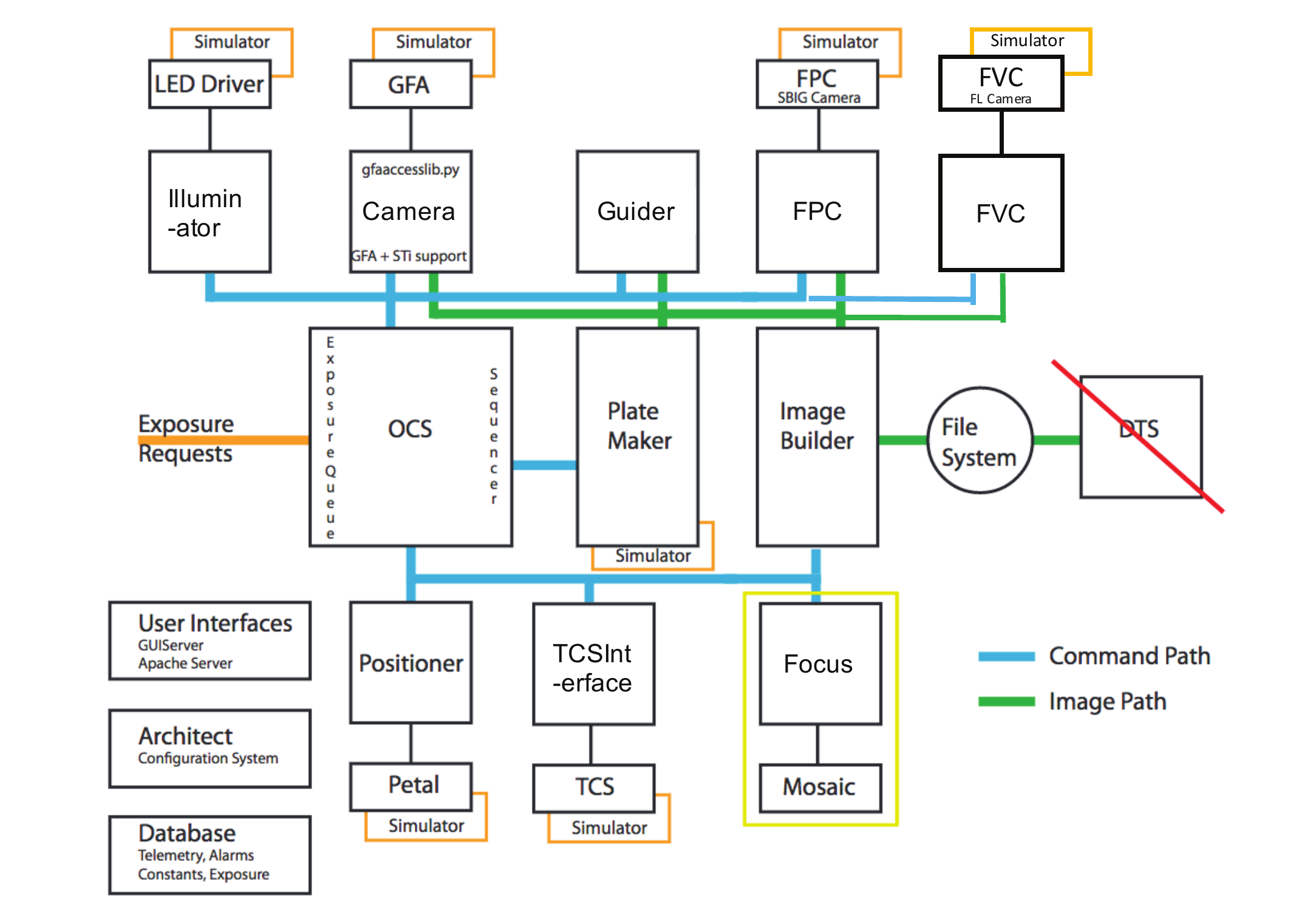}
                \caption{Schematic view of the ProtoDESI ICS. Exposure requests are submitted to the OCS application that coordinates operation of the instrument using dedicated applications for the illuminator, GFA camera, FPC, the positioners as well as an interface to the Mayall telescope control system (TCSInterface). Instrument focus is controlled via the focus application and the existing Mosaic corrector focus motor. Guide images are processed by the guider application and corrections signals are sent to the TCS via the TCSInterface. The OCS uses PlateMaker to transform the requested on-sky coordinates to either focal plane or FVC coordinates and images of the back illuminated fibers taken with the FVC are analyzed to confirm that all positioners have reached the requested target postions. After an exposure is taken, the Image Builder application collects data from the GFA and FPC cameras, combines it with telemetry and meta data information and writes everything as a multi-extension FITS file to disk. Note: the Data Transfer System (DTS) was not used on ProtoDESI.}
                \label{ICS}
            \end{figure*}

            The DESI ICS performs all control and monitoring functions required to operate DESI. One important aspect of the ProtoDESI project was early integration of several DESI hardware subsytems with the ICS prior to DESI commissioning. Wherever possible, ProtoDESI was run  with  elements of the  ICS similar to the expected final version.  This included all data acquisition and flow, connection to the Mayall TCS, monitoring infrastructure, operations database, guiding, PlateMaker (Sec. \ref{sec:platemaker}), subsystem coordination, and user interfaces (Fig. \ref{ICS}). Additionally, the computing hardware architecture, similar to the design for DESI commissioning, was used for the first time. This included three server class racks, a disk array, and two iMac observer consoles \citep{spie_ics}. 
        
            At the core of the ICS is the DESI Online System (DOS). This includes an application framework for the subsystems' distinct software. It is built on Pyro\footnote{https://pypi.python.org/pypi/Pyro4/}, an object-oriented communication package which enables efficient command and data transfer. DOS also contains all of the user interfaces, which display exposure sequences, real-time telemetry, and image previews. The Observation Control System (OCS), which orchestrates exposure sequences, is based on the architecture developed for the DECam on the Blanco telescope, which was successfully deployed in 2012 \citep{DES}. The OCS controls the flow of data from one subsystem's software to another using a sequence oriented architecture. The structure of the OCS was validated through a variety of routine sequences, including guiding, taking dark and flat images, and  a dedicated object exposure sequence customized for ProtoDESI that will be described in Section \ref{sec:expsequence}. DESI will collect a large amount of metadata and telemetry information, which will be recorded in the operations database. The operations database is implemented using PostgreSQL and ProtoDESI served as a testing ground for its architecture and user interfaces. The observers also had access to an electronic log book to keep a record of observation sessions, in particular any issues that came up during observing.
        
            \subsubsection{PlateMaker Software}
                \label{sec:platemaker}
                One of the key elements of the ICS is PlateMaker, responsible for making all critical coordinate transformations to place fibers on targets. ProtoDESI tested the data flow to and from PlateMaker, and its performance enabled us to meet our goals. During an exposure sequence, PlateMaker runs through several processes, including calculating apparent target positions, guide star identification, and calculating fiber positioner locations based on FVC images.
            
                The initial target positions are provided to PlateMaker in tangent plane coordinates, with the field center and a list of astrometric standards including potential guide stars. For ProtoDESI, astrometric standards and guide stars were selected from the NOMAD (Naval Observatory Merged Astrometric Dataset) catalog, and later the Gaia DR1 catalog \citep{Gaia}. PlateMaker applies offsets, rotations, scale and skew corrections (Eq. \ref{one}), then incorporates aberrations, refraction and polar axis misalignments to produce the final apparent targets (Eq. \ref{two}). The corrected target locations are output along sky North ($\xi$) and sky East ($\eta$) rather than RA/DEC:
                \begin{eqnarray}
                    \label{one}
                    \xi_0 = \left[ (y-y_0)\,\cos\theta_0 + (x-x_0)\sin\theta_0\right] \frac{sp}{3600} \nonumber \\
                    \eta_0 = \left[ -(x-x_0)\,\cos\theta_0 + (y-y_0)\sin\theta_0 \right] \frac{sp}{3600}
                \end{eqnarray}
                where $x_0$, $y_0$ are the locations of the telescope boresight, $\theta_0$ is the rotation of the GFA relative to the ProtoDESI focal plate, $s$ is the focal plane scale factor, and $p$ is the pixel size of the GFA (15 $\mu$m); 
                \begin{eqnarray}
                    \label{two}
                    \xi = \xi_0 (1+R+A+R\sin^{2}\psi \tan^{2}z) + \nonumber \\ 
                    \eta_0 R \cos\psi \sin\psi \tan^{2}z - \theta\eta_0 \nonumber \\
                    \eta = \eta_0 (1+R+A+R\cos^{2}\psi \tan^{2}z) + \nonumber \\
                    \xi_0 R \cos\psi \sin\psi \tan^{2}z + \theta\xi_0.
                \end{eqnarray}
                R and A are the factors for refraction and aberration, $\psi$ is the position angle of zenith, $z$ is the zenith angle, and $\theta$ is the position angle of the sky in the North direction relative to the focal plane sky North direction, including contributions from precession, polar axis misalignment, and overall rotation of the ProtoDESI focal plate.
            
                In order to derive focal plate coordinates from the sky coordinates $(\xi,\eta)$, one needs to know the distortion map of the corrector \citep{spie_platemaker}. In general, one can not assume that the telescope/corrector system is fully symmetrical, requiring a more general model than 1D polynomials. For DESI, these non-axisymmetric components are modeled as follows. The wavefront error $W$ can generally be written as a function of the exit pupil coordinates $(\rho,\psi)$ (Eq. \ref{zernikes}) using Zernike radial polynomials $R_l^s$:
                \begin{eqnarray}
                    \label{zernikes}
                    W(\rho,\psi; r,\theta) = \sum_{l}\,\sum_{s} [ A_{ls}\cos(\psi-\theta) \nonumber \\
                    + B_{ls}\sin(\psi-\theta)] R_{l}^{s}(\rho).
                \end{eqnarray}
                The coefficients $A_{ls}$ and $B_{ls}$ can be written as a function of position in the focal plane using one form of spin-weighted Zernike polynomials ${}_{s}^{*}Z_{n}^{m}$:
                \begin{equation}
                    (-1)^{s}(A_{ls}+iB_{ls}) = \sum_{n}\sum_{m}\,({}^{*}a\,_{nm}^{ls}\, - i\,{}^{*}b\,_{nm}^{ls})\,{}^{*}_{s}Z_{n}^{m},
                \end{equation}
                where
                \begin{equation}
                    {}^{*}_{s}Z_{n}^{m} = R^{m+s}_{n-s}(r)\, e^{im\theta}.
                \end{equation}
            
                In these equations, $-n\leq m\leq n-2s$, $n+m$ is even, and ${}^{*}a_{nm}^{ls}, {}^{*}b_{nm}^{ls}$ are complex and Hermitian on index $m$. Distortion corresponds to terms with $l=s=1$. For computational purposes, the complex summations are rewritten as a pair of real summations \citep{kentprep}. For DESI, 16 terms are needed (compared with 42 needed for a more traditional mapping). For ProtoDESI, 10 terms are needed. Initial values for the $a$ and $b$ coefficients are obtained by making fits to raytraces of the corrector optical design. The initial focal plate locations for the target fibers are calculated assuming that the telescope has been re-positioned to point precisely at the field center.
                
                With the initial acquisition images from the GFA, PlateMaker identifies stars based on pattern matching with the catalog. Knowing the apparent sky coordinates that correspond to the GFA pixels coordinate space, plus the known location of the GIFs relative to the GFA pixels, PlateMaker computes the apparent sky coordinates at the location of each GIF using the known optical distortion pattern to extrapolate the GFA astrometric solution to the GIFs.
            
                After the initial move, the fibers and fiducials are backlit, and the FVC images are taken to measure their locations. The GIF pinholes, having known astrometric coordinates, are used to calibrate the FVC pixel space. The true location of the fibers are determined from the FVC images and offsets from their targets are calculated. These are transformed back to the focal plate and interpreted as delta movements for the positioners; the process is repeated until at least 2 fibers are within the required 10 $\mu$m RMS error of their desired locations. At the same time, the FVC pixel coordinates of the backlit fiducials are compared with the lab-based metrology coordinates of each fiducial in the focal plane and used to generate updated coefficients in Eq. \ref{zernikes}. Images of the focal plane by the FVC and on-sky images from the GFA were required to initialize the PlateMaker software algorithms, after which it operated as an integrated application of the OCS. 

    \section{Integration and Commissioning}
    \label{sec:commish}
\begin{figure*}
                \centering
                \includegraphics[width=10cm]{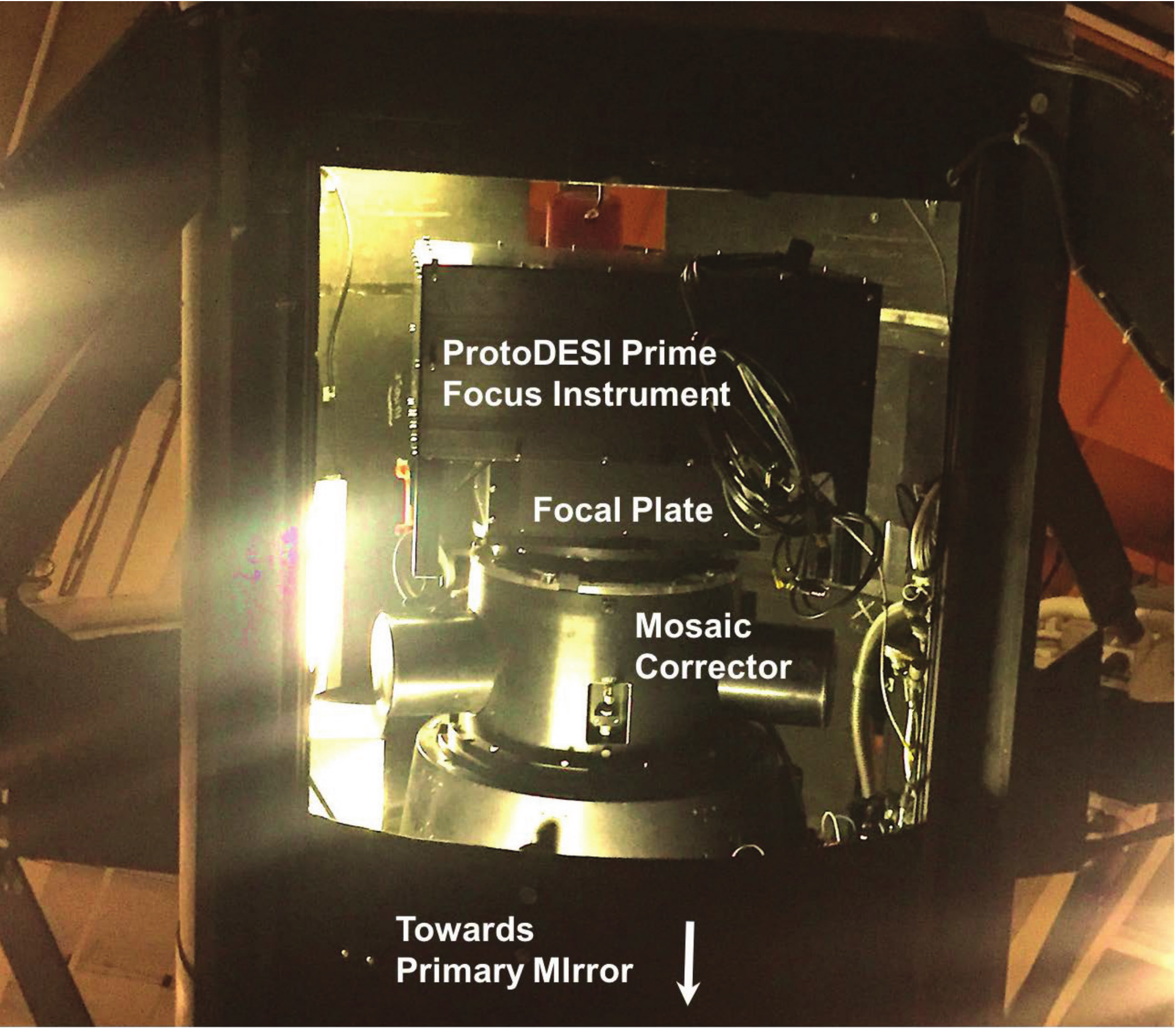}
                \caption{The ProtoDESI prime focus instrument as mounted behind the Mosaic corrector after installation. The bottom of this image is in the direction of the primary mirror.}
                \label{install2}
            \end{figure*}

    	The ProtoDESI prime focus instrument was assembled at LBNL, where functional capabilities were tested and the performance was benchmarked prior to shipment to KPNO. The ProtoDESI focal plate instrument arrived at Kitt Peak on Aug. 14, and after a quick ($\approx$1 hr.) installation (Fig. \ref{install2}), a functional test was performed that confirmed that all subsystems had survived shipment, received adequate power and could be commanded with an external control system via ethernet.  Prior to this, the FVC was shipped to KPNO, integrated with the 600 mm lens and mounted in the Cassegrain cage of the telescope. The ICS hardware was configured and all software installed so that testing could begin with the arrival of the prime focus instrument. The fully assembled prime focus instrument was shipped as one piece requiring minimal preparation when it arrived at Kitt Peak.
        
        Our commissioning campaign included subsystem tests that ocurred both in the lab and integrated on the telescope. These tests characterized the performance of the subsytems and determined their compliance with requirements. The commissioning test sequence increased in complexity with time, beginning with tests that used stand alone software that could be done in the lab and then with the dome closed. These included positioner accuracy, FPC and FVC tests. We then moved to on-sky testing for the GFA, the first successful image taken on Aug. 25. When these tests were complete, the ProtoDESI instrument was sufficiently integrated and characterized to attempt guiding and aligning fibers with targets, at which time we moved into the operations phase (Sec. \ref{ops}). Below we describe the tests and results that comprised the commissioning campaign, organized by subsystem. 
        
	        \subsection{FVC Centroiding}
            \label{sec:fvc_centroid}
            The purpose of the FVC is to provide a feedback mechanism to PlateMaker and the robotic positioners, helping ensure that the positioners are centered on targets. To achieve this, the FVC is required to deliver centroid locations of the fiducials and backlit fibers to 1/30 of a pixel, corresponding to a precision of less than 3 $\mu$m on the focal plate. While the FVC was used in the lab as discussed in the next section, characterization of the true FVC performance had to be completed on the telescope due to the separation from the focal plate and presence of corrector optics. To test the accuracy and stability of the FVC, a dataset was collected at different times of day with a variety of dome conditions and exposure times. The LEDs that backlit the fibers and the fiducial brightness were adjusted so that the flux on the FVC at all exposure times was more or less the same. Using the large dataset collected, we first evaluated the precision of our measurement of the fiducial locations (defined as the center of the fiducial). To compare images with the same exposure time to one another, we used a linear transformation to put all exposures into the same coordinate system, since the position of the camera relative to the focal plane could change. Taking the mean error of all fiducial measurements vs. exposure time it was found that the precision requirement could be met for the fiducials with exposure times as low as 0.5 seconds. 
            
                \begin{figure*}
                    \centering
                    \includegraphics[width=10cm]{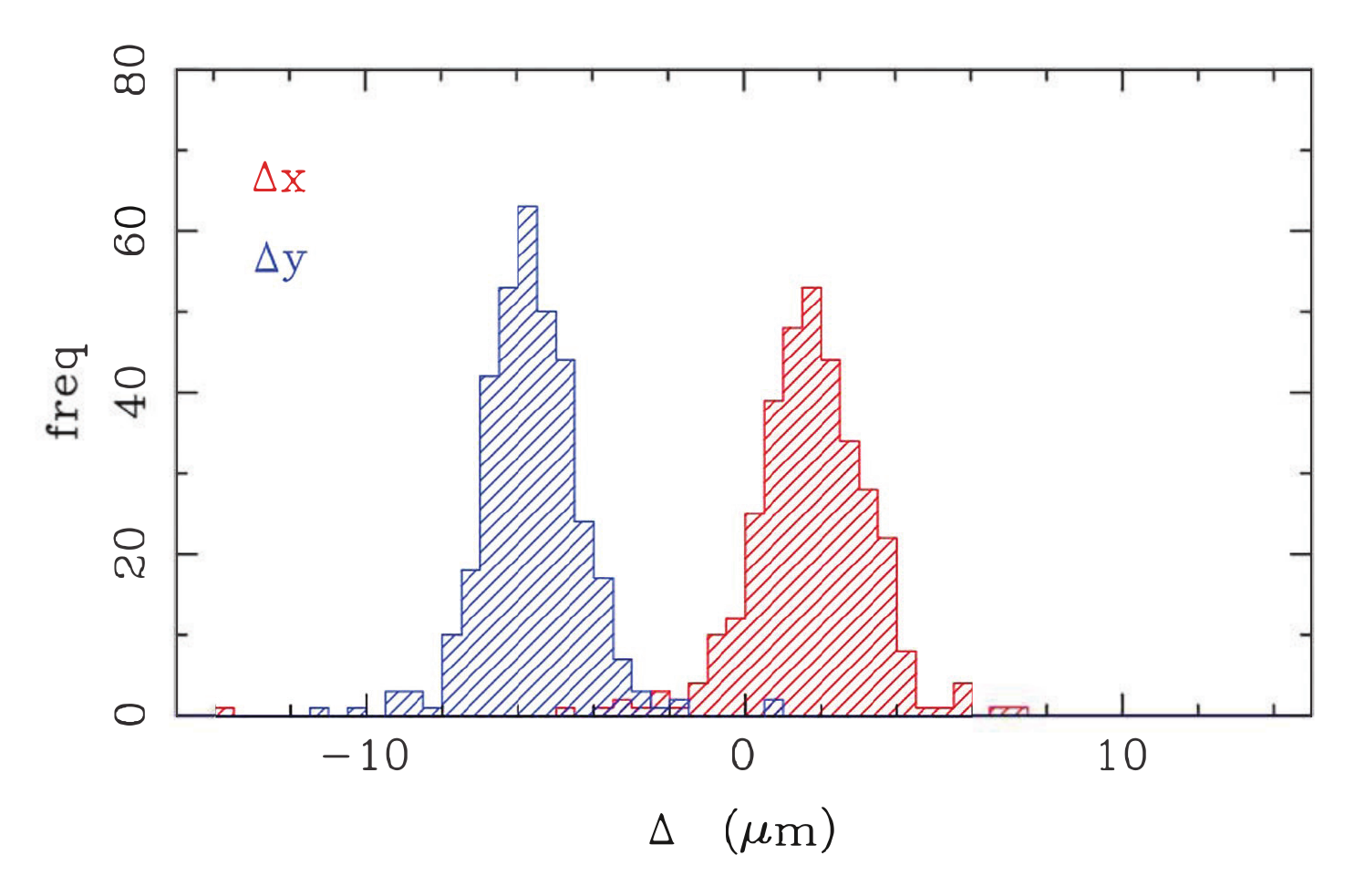}%
                    \includegraphics[width=6cm]{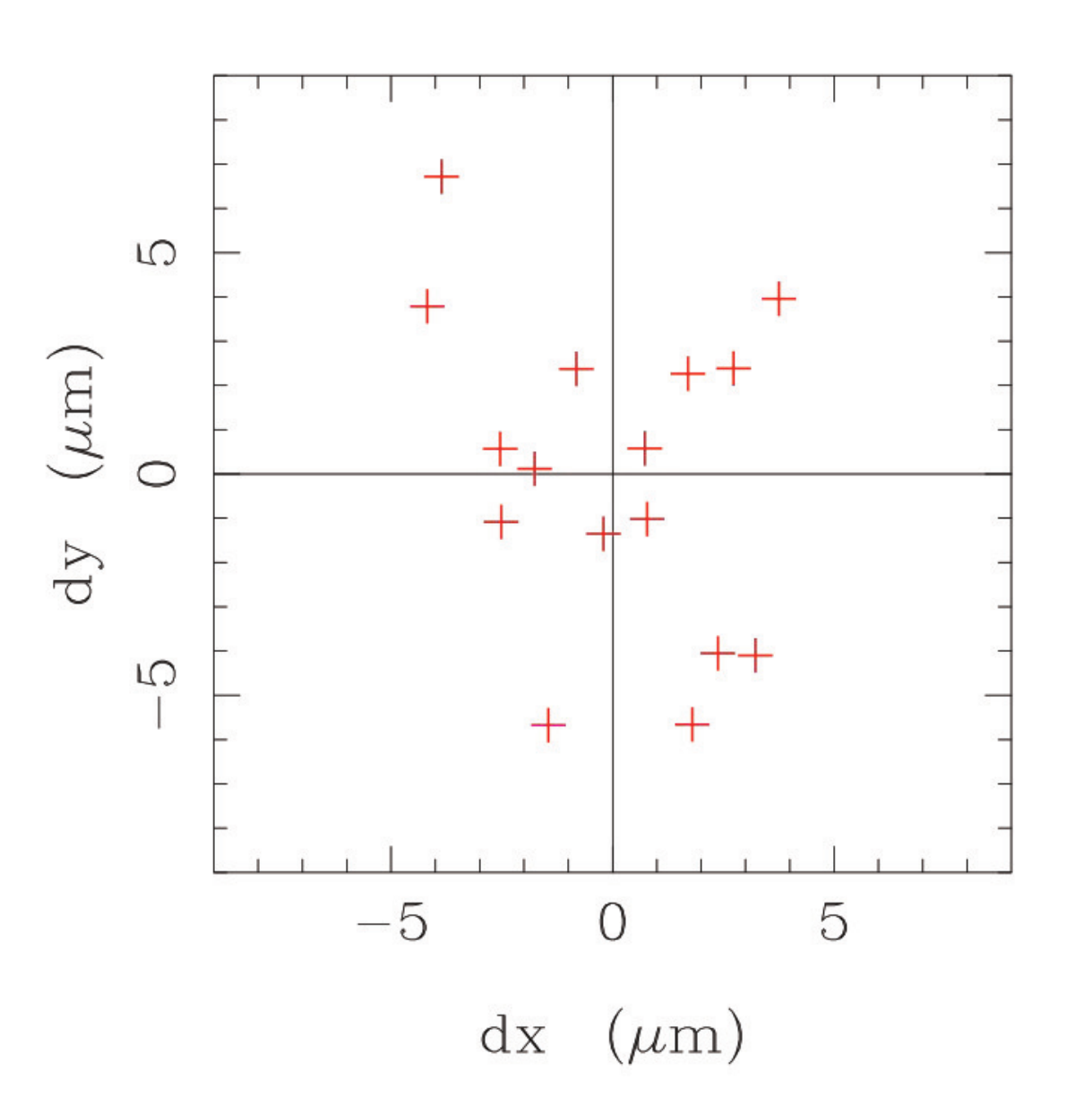}
                     \caption{\textbf{Left:} Difference between the metrologically measured location of a single fiducial and its location measured by the FVC. These results are consistent with the 5 $\mu$m precision of the CMM used. \textbf{Right:} The average FVC offset of all 16 fiducials from their metrologically measured position.}
                    \label{fidoffsets}
                \end{figure*}
                
\begin{figure}

                    \includegraphics[width=8cm]{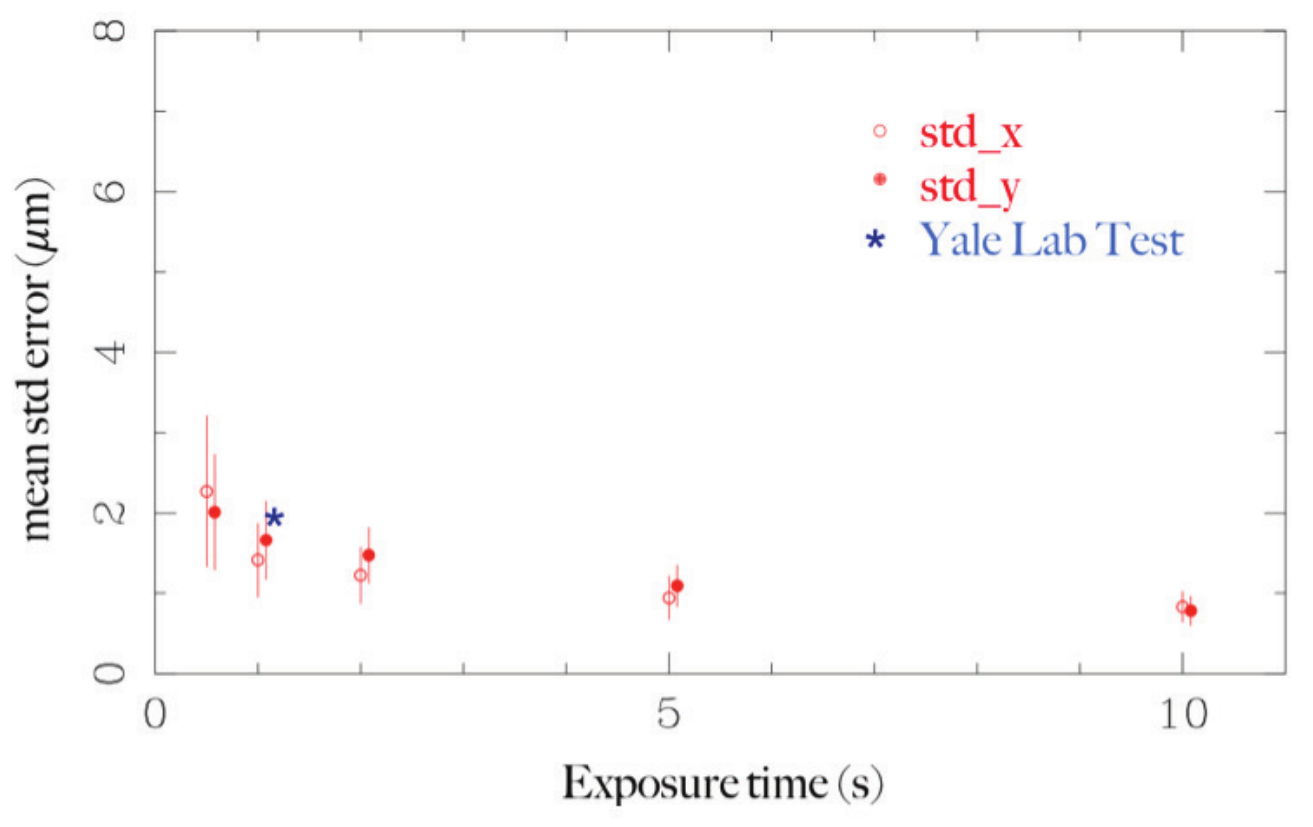}
                    \caption{The standard error on the FVC centroid measurements of the fibers after correcting for  distortions. Each point represents the mean of 70 images. The blue asterisk represents the in-lab results prior to delivery.}
                    \label{fiber_rms}
                \end{figure}

                This analysis was insensitive to distortions of the Mayall corrector lens. With the dataset of $\approx$450 exposures, we were able to solve for the distortions to third order. With this distortion calculation, each image could be pre-corrected to get an even better calculation of the average deviations of the fiducial locations. It was found that all the fiducials were offset in a random distribution from the CMM metrology measurements made at LBNL (Fig. \ref{fidoffsets}), consistent with the $\pm$5 $\mu$m precision of the CMM used. The distortion map derived from the FVC images were consistent with that used by PlateMaker in Section \ref{sec:platemaker}. Using this distortion map, the positions of the fibers could be accurately measured. The results (Fig. \ref{fiber_rms}) indicate that the centroids of the fibers can be measured with the same precision as the fiducials and meet requirements at all exposure times of 0.5 seconds and above (the design exposure time was 2 seconds). During DESI, more of the FVC FOV will be used so additional distortion errors are likely to be encountered. However, there will be many more fiducials (100 FIFs, 20 GIFs) to enable additional constraints. 
                
        \subsection{Positioner Accuracy Tests}
        \label{sec:acctest} 
\begin{figure*}
                \centering
                \includegraphics[width=11cm]{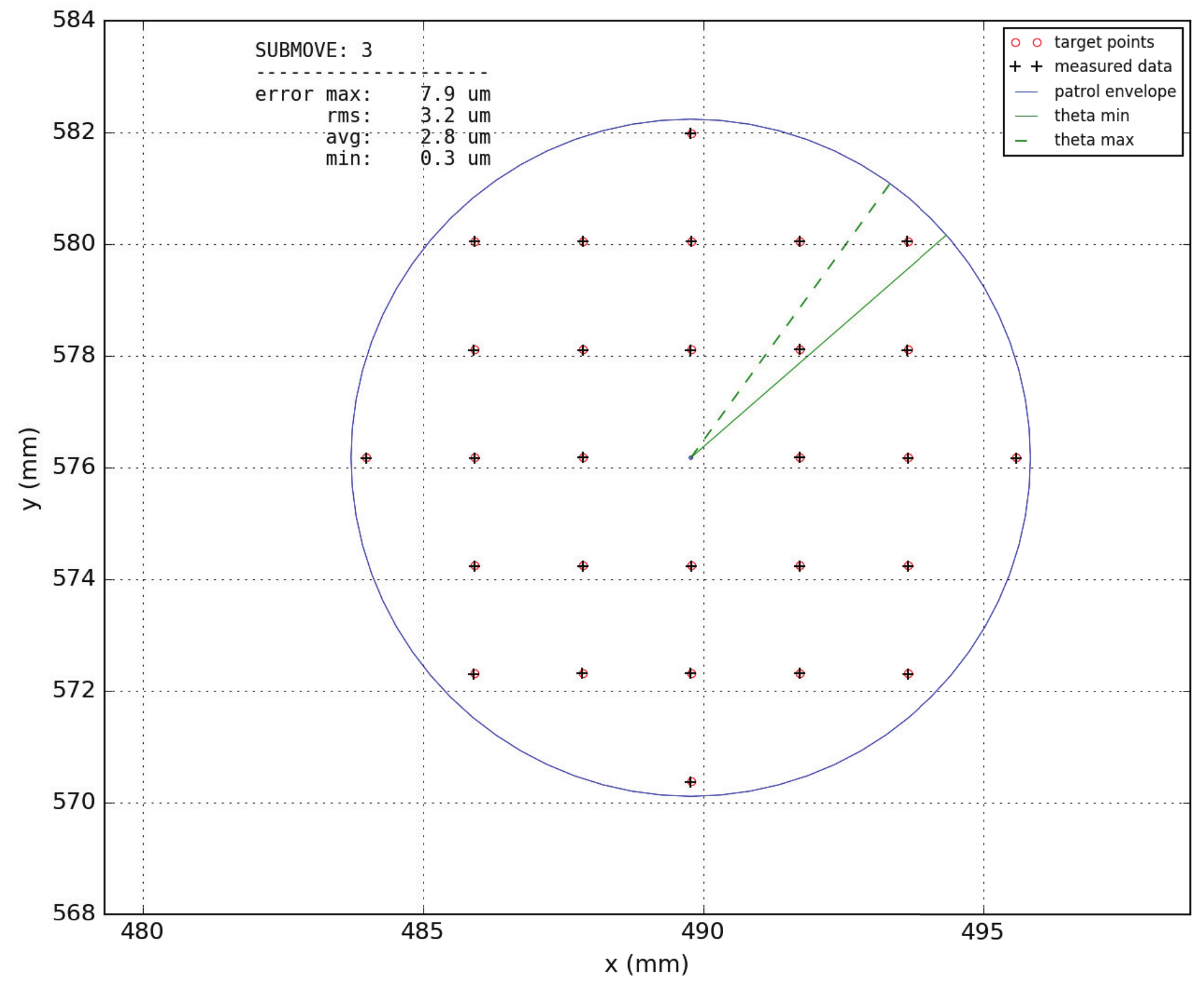}
                \caption{Output of a positioner accuracy test performed at LBNL on a positioner installed on ProtoDESI, which is identical to those performed at Kitt Peak. This test was performed on Aug. 8, 2016 and contains a grid of 28 positions within the patrol area. Each point on the plot represents a location within the patrol area which is targeted up to four times with increasing accuracy. After three corrective submoves, the positioner meets the requirements with an RMS error of 3.2 $\mu$m. }
                \label{xy_acc_test}
            \end{figure*}

        In order to test the performance of the positioners it is necessary to  use a fiber view camera, without which it is impossible to determine the true location of the fibers. Before running the accuracy tests it is necessary to run a positioner calibration sequence. The robots are commanded to move blindly to a series of pre-defined points. FVC images are taken at each point, and the measured positions are used to calculate 6 calibration parameters: kinematic radius of theta and phi axes, angular zero points of each axis, and the (x,y) center of the robot in focal plane coordinates. In accuracy tests, all positioners are also moved blindly on a grid of pre-selected points. The locations of the fibers are measured by the FVC and sent back to the positioner software where a delta correction move is calculated. Each positioner performed corrective moves (submoves) 3-4 times attempting to get closer to the target position, with an FVC measurement after every move. When the test was complete, the maximum, minimum, and RMS error were calculated for each submove (Fig. \ref{xy_acc_test}).
        
        The positioners were designed to arrive at their commanded position with an RMS error of less than 5~$\mu$m with $\leq$3 corrective moves, ensuring an on-sky RMS accuracy of $\leq$10 $\mu$m. While additional corrective moves may reduce RMS error, during DESI operations reconfiguration of all positioners is expected to be complete to the required accuracy within 45 seconds. The initial blind move of the positioners, with no feedback from the FVC, is required to arrive within 100 $\mu$m of the commanded position. These requirements are expected to be met under all possible operating conditions for DESI.
        
        The three ProtoDESI positioners were tested both at the University of Michigan and at LBNL prior to shipment to Kitt Peak.  We did not have the laboratory space to use the FVC with a 600 mm lens, but we used a similar Canon lens with a focal length of 100 mm, placed it $\approx$2 meters from the focal plate so that when imaging the fibers they had the same demagnification of $\approx$22 as expected on the telescope. After adjusting the fiducial brightnesses to that of the backlit fibers, we imaged the front of the focal plate with the FVC. The initial tests of the positioners at University of Michigan indicated that their performance was excellent, exceeding the requirements. When the positioners were integrated in the ProtoDESI focal plate at LBNL, the performance was slightly diminished, likely due to the test configuration and conditions (e.g., stray light, floor vibrations), but all positioners still met the requirements.
             
             The first tests conducted after completion of the initial functional tests on the telescope were the positioner accuracy tests. When mounted on the telescope the performance decreased significantly compared to results from lab testing. At the third submove, the RMS error was $\approx$20 $\mu$m with the dome closed, decreasing to $\approx$10-14 $\mu$m when the dome was opened. This change in performance on the telescope corresponded to dome seeing effects.
        
            \begin{deluxetable*}{ccc}
                \tablecaption{Positioner Accuracy Test Results \label{table:acc} }
                \tablecolumns{3}
                \tablewidth{0pt}
                \tablehead{
                \colhead{Environment} &
                \colhead{Blind Move Max Error ($\mu$m) } &
                \colhead{Submove 3 RMS error ($\mu$m)} 
                }
                \startdata
                    Requirement & 100 & 10\\
                    UM, pre-shipment & 22 & 0.4\\
                    LBNL, integrated with ProtoDESI & 35 & 3 \\
                    Telescope, dome closed, no PlateMaker & 60 & 25\\
                    Telescope, dome opened, no PlateMaker & 50 & 14\\
                    Telescope, dome opened, PlateMaker & $\approx$50 & 4-6 \\
                \enddata
            \end{deluxetable*}
        
            After identifying the dome seeing issue, tests were run with an open dome and sufficient circulation, but the accuracy requirements were still not met. The software used for these early commissioning tests was not fully integrated in ICS, as the test did not require the OCS and image building. Because of this, the transformations from FVC centroid measurements to focal plate coordinates was completed using a linear scaling by the stand alone positioner software, which proved to be an inadequate model of the corrector's optical transfer function. On September 18, we integrated a subset of the PlateMaker algorithms into the accuracy test software, correctly translating the FVC centroids to \textit{x-y} locations on the ProtoDESI focal plate.  With this additional functionality, the RMS error finally dipped below 5 $\mu$m and continuously met the requirement of $\leq$10 $\mu$m RMS error (Table \ref{table:acc}). Once we were meeting the requirement, we did so within 1-2 corrective submoves. During this testing campaign, it was determined that the positioner calibration sequence needed to be run at least once a night. 
                    
            \subsection{FPC Performance}  
            \label{sec:fpc_results}
                The requirement on the photometry camera was $\approx$1\% relative photometry error for each fiber, corresponding to a few microns error on the measurement of the centroid location of the star within the fiber. While in the lab, we did not project targets on the robotic positioners, but created a diffuse light source that could illuminate all three positioners, simultaneously measuring the output of the fibers as the positioners moved. Preliminary data analysis of the FPC images was performed and the camera was characterized, indicating that we would meet the requirement. On the telescope, using the calibration screen in the telescope dome, we took several series of 11 flat-field images with the FPC, each at a different exposure time. For each exposure time, the output from each fiber from the flat illumination remained constant with $\leq$0.8\% error, meeting our measurement goals.
                
\begin{figure}
                    \centering
                    \includegraphics[height=4.5cm]{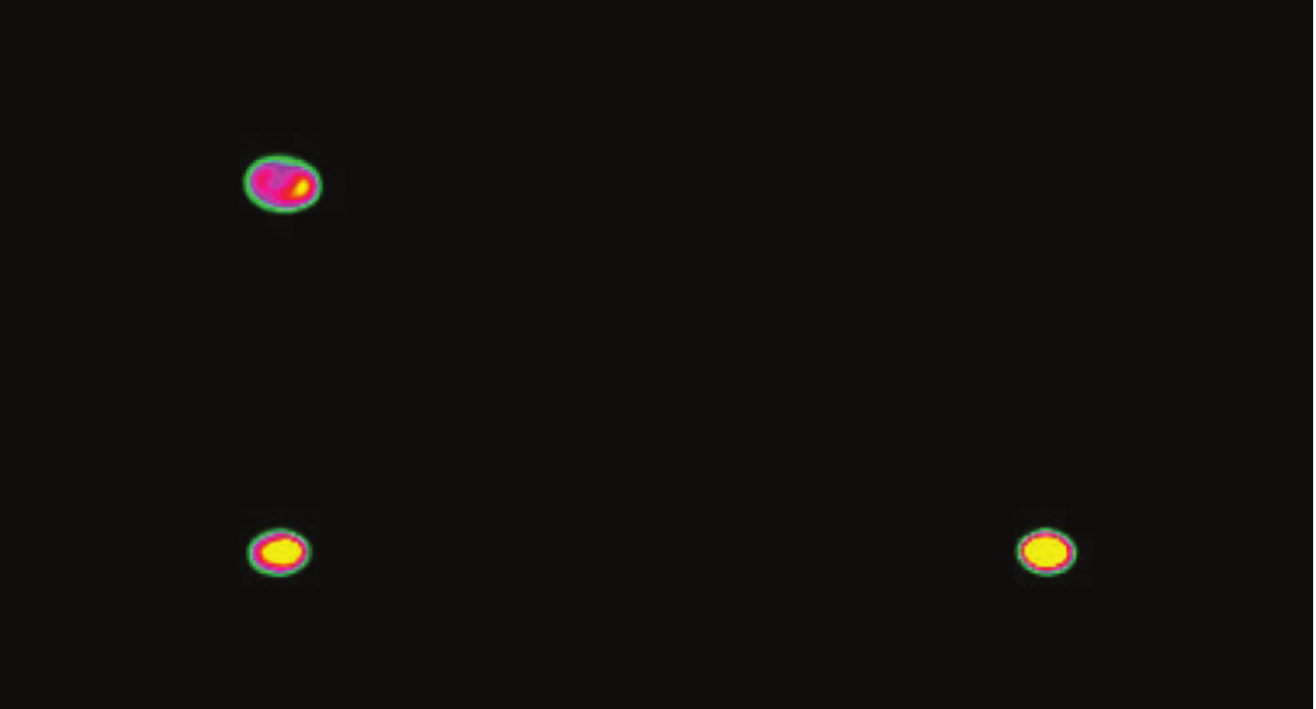}
                    \caption{Output of the three fibers during a 10 second flat-field image taken with the FPC. The upper left fiber has clear structure in the output.}
                    \label{3fibs}
                \end{figure}

\begin{figure}
                    \includegraphics[width=9cm]{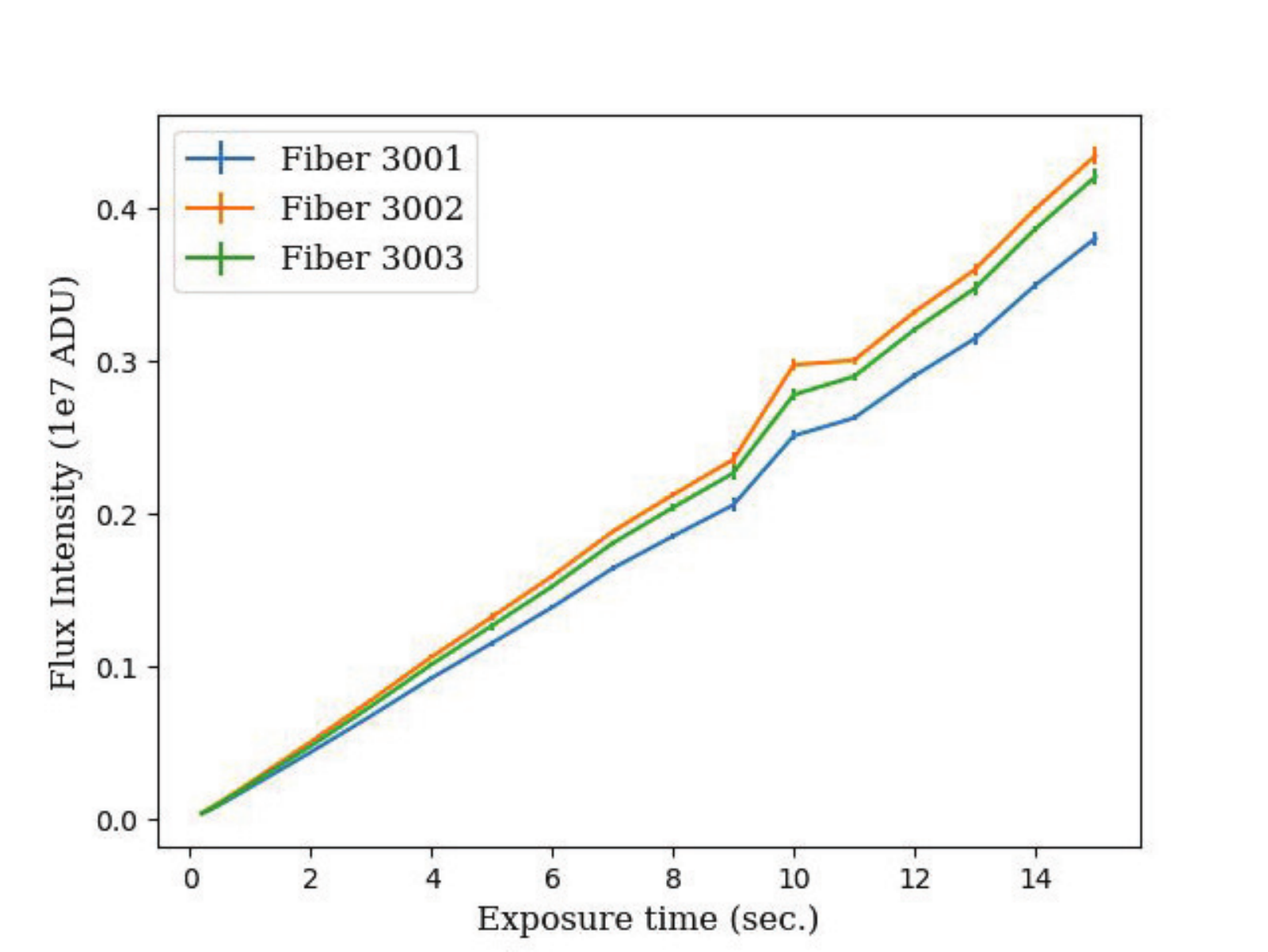}
                    \caption{Relative throughput and linearity of each fiber. These measurements were made using dome flats after the prime focus instrument was installed on the telescope. Data with 10 s exposure time were taken at a different CCD temperature, hence the outliers. Fiber \#3001 and \#3003 had 87.0\% and 95.5\% respectively of the throughput of \#3002, the best performing fiber.}
                    \label{linearity}
                \end{figure}

                Since our objective was to measure the relative photometry, overall throughput was not a priority; however, care was taken to handle the fibers and maintain the minimum bending radius. Despite this care, one of the fibers appeared to be offset in the fiber array block such that the output did show some structure (Fig. \ref{3fibs}). While each fiber did have a different throughput, when aligned with a star, they all had sufficient signal to noise with a FWHM of $\approx$25 pixels. The throughput differences in these fibers are not a concern for DESI as the routing and termination will be quite different. The linearity of the camera was measured in the lab (Fig. \ref{linearity}) and then again on the mountain, with comparable results. Additional analysis of the FPC images taken on the telescope indicated that the location of the center of the fiber images moved up to 20 pixels from image to image, likely due to a shift of the camera relative to the fiber array block for different telescope pointings. 
            
\begin{figure}
                    \centering
                    \includegraphics[width=9cm]{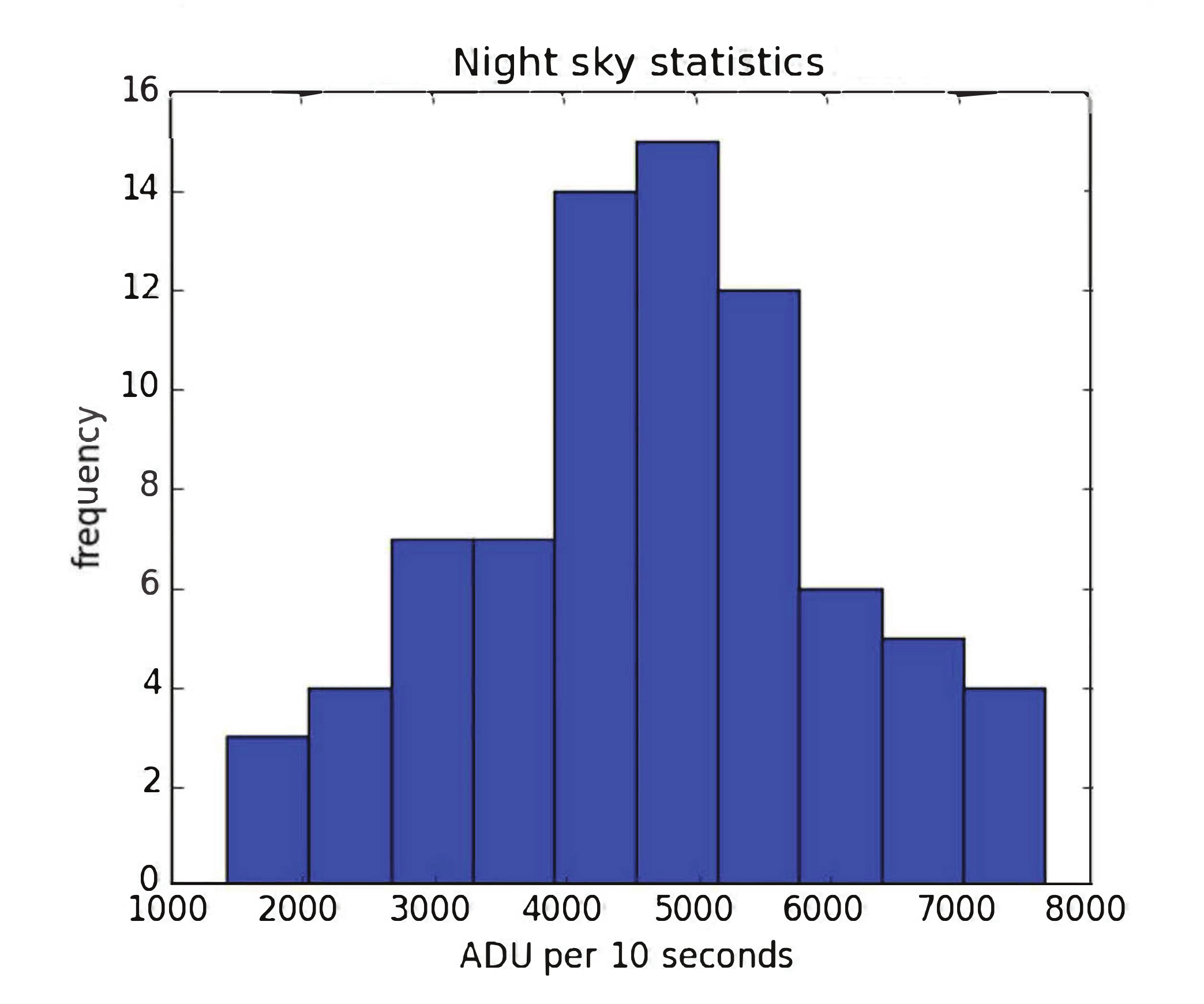}
                    \caption{Histogram of 77 night sky measurements from a single fiber over a few independent FPC image runs. The best fit maximum, assuming a Gaussian distribution, is 4500 ADU, which is consistent with the expected sky brightness in the r-band at Kitt Peak. However the breadth of the distribution, 1500 ADU, was about twice its expected breadth.}
                    \label{night}
                \end{figure}

                There were several fields that only had targets for one fiber at a time, while the other fibers were pointed at the sky background. This gave us a unique opportunity to look at the brightness of the night sky, helping to establish the signal to noise of the camera. We measured the signal from the night sky to be $\approx$4500 ADU for a 10 second exposure (Fig. \ref{night}), consistent with 20.6 mag/arcsecond$^2$ in the r-band as listed in the KPNO Direct Imaging Manual\footnote{https://www.noao.edu/kpno/manuals/dim/} appendix. The noise from these sky measurements, nevertheless, was much larger than expected. While we expected a SNR of 6 we measured closer to half that, which indicates that the DC voltage camera offset value of the FPC was not entirely steady but shifted the signal by variable amounts.
            
        
            \subsection{GFA testing}
            \label{sec:gfa_testing}
            While the GFA cameras on DESI will be responsible for guiding, focus and alignment, the ProtoDESI GFA was only used for guiding and alignment. It was designed to measure centroids of guide stars with 30 milliarcsecond (mas) accuracy at 1 Hz. The GFA used on ProtoDESI was a prototype, and our experience with it was extremely useful in improving the final design for DESI. The GFA was the final subsystem to be integrated with the focal plate at LBNL. Before shipping the prime focus instrument to Kitt Peak, there was only time to complete functional tests, confirming that the CCD sensitivity was sufficient to reconstruct an image with adequate SNR. There was not enough time to fully characterize the camera and optimize its parameters prior to installation on the telescope, but a preliminary gain measurement was made in the lab. The initial results indicated that the ADC was using only a small dynamic range available and that the system was non-linear due to differences in the four readout amplifiers. 

              During integration with the focal plate, it was discovered that there was a loose connection in the Kapton flex cable which ran from the GFA to its power and ethernet connectors. This was addressed before shipment, but on the mountain it was found that the problem persisted and when it caused the GFA to shut down, it required opening the prime focus instrument and manually rebooting the camera. Towards the end of the ProtoDESI campaign, the firmware was modified, enabling reboot via software. The origin of the GFA shutdowns was not explicitly identified, but we postulate there were two causes. First, during slew of the telescope, the loose connection of the flex cable could temporarily break the power connection. More commonly, however, the interlock system installed on the power sources (see Sec. \ref{sec:pf}) would shut down if the temperature of the GFA's FPGA exceeded $60\degree$C. The cooling mechanism for the camera was a fan that pulled air from behind the CCD to outside the focal plane, and while this was expected to run continuously, the fan sometimes shut off. If the fan was turned off for more than $\approx$1 minute, the GFA would approach the maximum allowed temperature.
                
              After installation on the telescope, while the GFA was powered on, significant pattern noise was seen in the images making guiding impossible. Initially this issue was assumed to be with the ethernet interface and was temporarily fixed by replacing the ethernet switch. It was eventually determined that there was a design error in the camera electronics that caused interface issues. With this resolved, the GFA delivered images with sufficient SNR to meet its requirements for guiding (see Sec. \ref{sec:guiding}). Gain measurements made on the telescope were found to be inconsistent with that measured in the lab, however, some other detector parameters were determined from on-sky images while ProtoDESI was installed. Using images of the trumpler37 field on Sept. 16 and 17 and assuming a gain of 4.5 e$^{-}$/ADU, a value that was not verified, the throughput was measured to be 0.46 combining the atmosphere, telescope reflectance, CCD QE, corrector and filter throughput. The read noise was measured to be 20 ADU/pixel and the dark current was $\approx$80 ADU/pix/sec at an ambient temperature of $\approx$20$\degree$C, dominating the sky noise. 

        \section{Operations}
        \label{ops}
            
            Once all subsystems were confirmed to function as expected, ProtoDESI moved on to its main goals: align a fiber with a target and maintain alignment for the duration of a DESI exposure. Prior to attempting the sequence of events that would align fibers with targets, we evaluated our guiding capabilities. The first successful object exposure sequence was accomplished on Sept. 14, 2016 when fibers illuminated by light from three stars were imaged by the FPC. After achieving acquisition of targets with the fibers, we extended the exposure sequence to include additional steps to measure the instrument's performance. These included telescope dithers to sample the pointing precision and alignment stability tests. Our goal was to build a dataset that would allow us to confirm the precision and stability to 5 $\mu$m, but due to non-ideal weather, we were only able to complete a subset of our planned tests. This section describes our operations approach. The results from all guiding, pointing precision, and stability tests are discussed in Section \ref{sec:results}.
        
                \subsection{Object Exposure Sequence}
                \label{sec:expsequence}
                    
                    Prior to a night of observing, we identified a number of fields that had enough bright stars for guiding and stars in the patrol areas of the three positioners to serve as fiber targets. From the list of available fields, one was selected based on time of the night and weather conditions, and the RA/DEC coordinates of the field center were given to the telescope operator. Once the telescope was in place, telescope control was switched to the OCS and the field information was uploaded, including the locations of the guide star(s) and fiber targets. While the OCS was capable of directly commanding the TCS to the field, for safety reasons, any telescope slew greater than 5 degrees required manual authorization for operation by the telescope operator. We then moved the prime focus stage in 50 $\mu$m steps, using GFA images at each step to find the best focus by analyzing how the PSF (Point Spread Function) of bright stars changed. The exposure sequence began with a full frame GFA image which PlateMaker analyzed to identify the guide star(s). With a guide star selected, the OCS entered guide mode, sending the ROI (Region of Interest) of the GFA image around the guide star to the guide software and forwarding guider corrections to the TCS. After entering guide mode, fiducials were turned on, fibers were backlit, and positioners made their first blind move. 
                    Passing GFA and FVC images through the PlateMaker software, as described in Section \ref{sec:platemaker}, the fibers were aligned with the targeted locations. At this point the fiducials and backlighting LED were turned off and a 10-second image was taken with the FPC, completing the sequence.
        
                \subsection{Dither Patterns and Stability Measurements}
                    
                    After achieving the primary goal of ProtoDESI, namely acquiring targets by aligning fibers with stars and maintaining pointing, the remaining two weeks of operations were spent running tests to evaluate the performance of our pointing and stability. To determine the stability, we took a 10 second FPC image approximately twice a minute for at least 20 minutes. The pointing accuracy of our system was measured by moving either the telescope or positioners in a $5\times 5$ dither pattern in an attempt to maximize the flux through the fibers. Assuming the true target location is where the flux through the fiber is at a maximum, we moved the telescope in either 1 or 2 arcsecond steps in a grid around the initial acquisition position of the fiber. The two step sizes were chosen so that we could identify an approximate maximum in a reasonable ($<$30 minute) amount of time. The larger step size of 2 arcseconds is just slightly bigger than the angular extent of the fiber so that the flux incident area for each step was independent and did not overlap. We also performed some dither patterns with the positioners, which could be dithered with much smaller step sizes. Unfortunately, the telescope was dithered in steps of RA/DEC and those of the positioners were \textit{x-y} positions on the focal plane, so the dither patterns could not easily be compared. The next version of PlateMaker will allow DESI to compare positioner and telescope dithers. The results from these tests are discussed in Section \ref{sec:pointing}.
        
                \subsection{Guide Modes}
                
                    During the object exposure sequence, guide stars were primarily selected by PlateMaker based on astrometry, but the guiding software could also be controlled directly by the observer. There were three guiding modes readily available for ProtoDESI: self, direct, and catalog. Before running the full object exposure sequence, we tested our guiding capabilities in self mode, in which the guiding software chooses a guide star from a full frame GFA image. The star selected had the highest SNR among those meeting several requirements, including: isolation within a 16-pixel radius, having a round PSF, located away from the edge of the image, and not being blended, saturated, or flagged for some other reason. The direct mode used a guide star pre-selected by the operator, identified in GFA pixel space. In this direct mode we were able to explicitly test our ability to measure guide signals on stars over a range of brightness (see Sec. \ref{sec:guiding}). 
                
\begin{figure*}
                        \centering
                        \includegraphics[width=12cm]{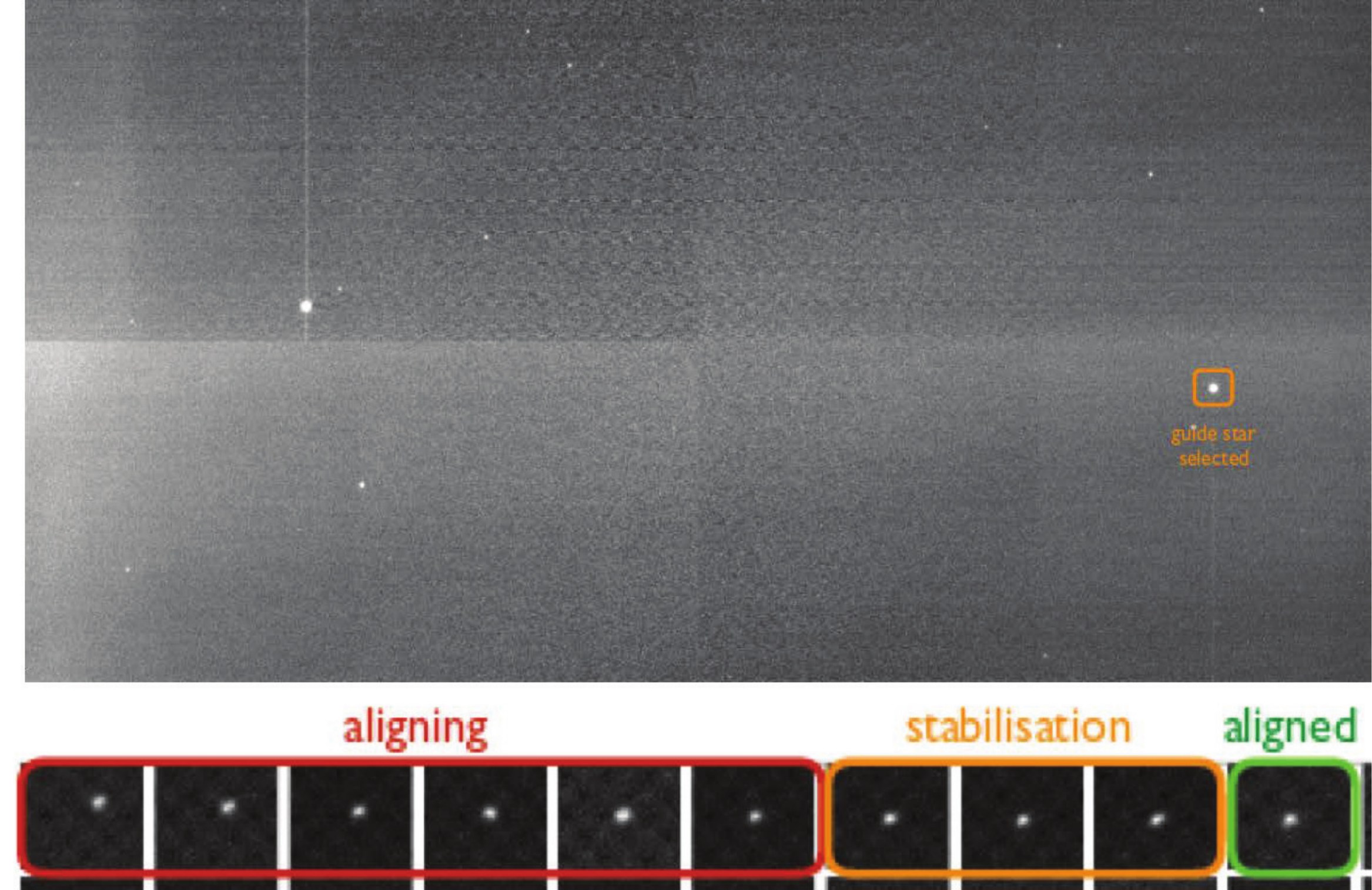}
                        \caption{\textbf{Top:} Full frame GFA image with the guide star identified. \textbf{Below:} The guider moves into ROI mode and sends correction signal to the TCS to align and stabilize the guide star.}
                        \label{guider}
                    \end{figure*}

                    When we moved to full exposure sequences, the catalog mode was used, in which PlateMaker was responsible for selecting the guide star. Once the telescope slewed into place, a full frame image was taken to confirm the current position. PlateMaker could then identify the guide star and select a pixel on the GFA for alignment. At this point, the guider reconfigured the GFA to operate in ROI mode, after which only postage stamp images of the guide star were received. In order to move the guide star to the selected pixel with the required accuracy, correction moves were sent to the TCS every few seconds. After achieving stability, the guider prompted the OCS to continue with the object exposure sequence (Fig. \ref{guider}).

                \subsection{Target Selection}
                
\begin{figure*}
                        \centering
                        \includegraphics[width=10cm]{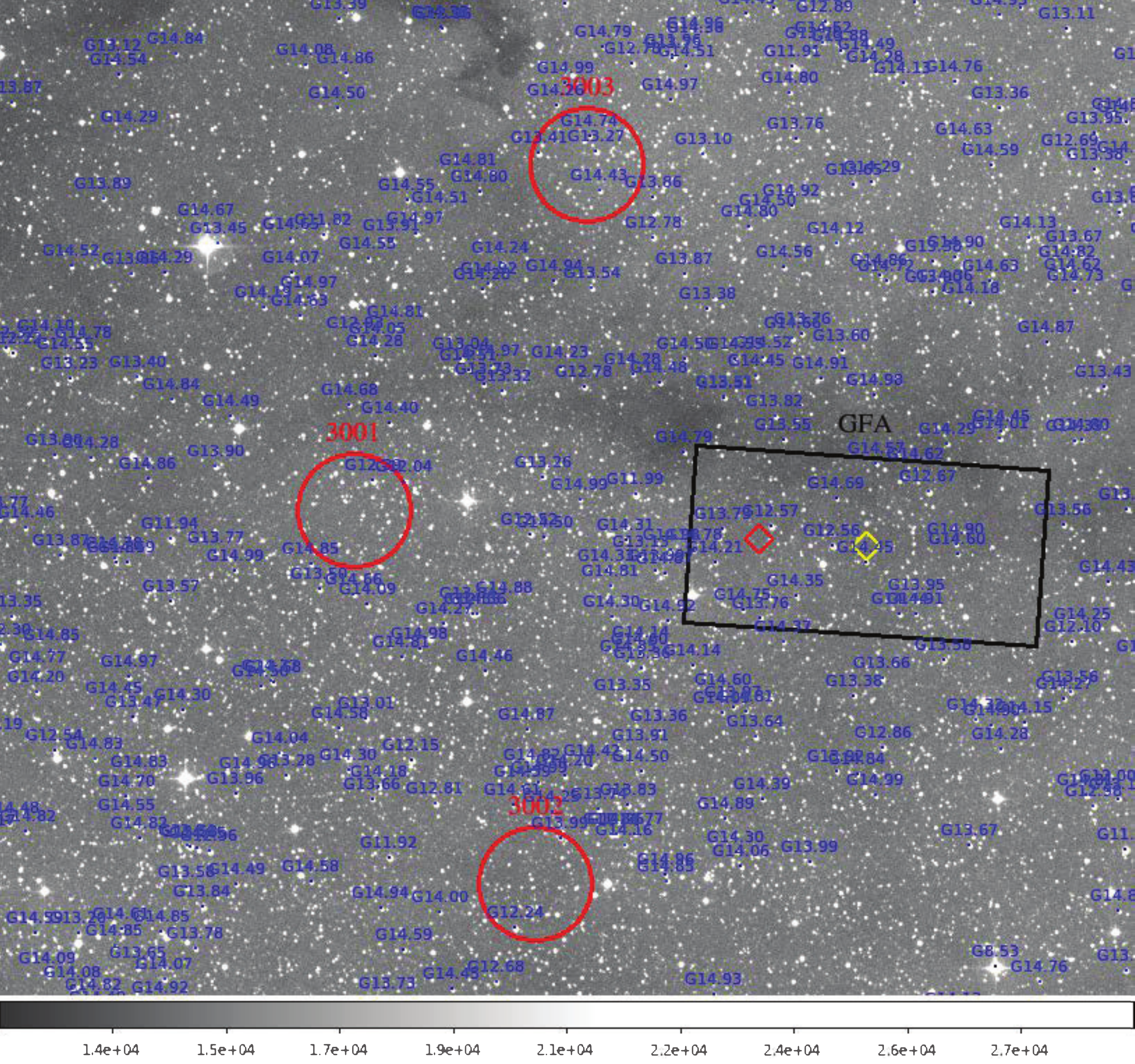}
                        \caption{Trumpler37 field, used on ProtoDESI to test pointing accuracy and guiding. The red circles represent the patrol area of each fiber and the black box is the GFA FOV.}
                        \label{trumpler}
                    \end{figure*}

                    The initial selection of ProtoDESI target fields was based on their accessibility given the observing dates, and designed to cover a range of sky positions and observing airmass. The fields chosen each included a few guide star candidates ($r_{\mathrm{AB}} < 17$) near stars separated such that they could be reached with the fiber positioners. The selected targets were bright ($8 \leq r_{\mathrm{AB}} \leq 12$) for pointing accuracy measurements. Our target files included the coordinates of the field center and also the coordinates of the three fiber targets. For some target files, we offset the positioners so that only one would be aligned with a star and the other two not used. The coordinates were first identified using the NOMAD catalog, but then moved to the Gaia DR1 catalog which had much better precision astrometry. All fields attempted are listed in Table \ref{table:targets}. We re-observed targets from night to night to trace repeatability. Figure \ref{trumpler} shows an example field, trumpler37, which we used to test both pointing accuracy and guiding, as it contained stars with a range of brightnesses in the GFA FOV.
                
                    \begin{deluxetable*}{ccccc}[h!]
                        \tablecaption{Target List  \label{table:targets}}          
                        \tablecolumns{5}
                        \tablehead{
                        \colhead{Name\tablenotemark{a}} &
                        \colhead{\# attempts\tablenotemark{b}} & 
                        \colhead{RA\tablenotemark{c}} &
                        \colhead{DEC\tablenotemark{c}} & 
                        \colhead{Catalog\tablenotemark{d}}
                        }
                        \startdata
                            hour0		& 47 &   2.4206 & 28.6358 & NOMAD\\
                            hour18\_5	& 55 & 277.1659 & 32.9500 & NOMAD\\
                            hour21 		& 86 & 314.3166 & 31.2828 & NOMAD\\
                            hour3 		& 25 &  43.9392 & 30.8636 & NOMAD\\
                            trumpler37 	& 50 & 324.4944 & 57.5317 & NOMAD\\
                            n7789a 		& 7  & 359.5963 & 56.7387 & NOMAD\\
                            33001 		& 4  & 276.4567 & 29.8505 & NOMAD\\
                            33002 		& 5  & 276.5449 & 30.0056 & NOMAD\\
                            33003 		& 3  & 276.5676 & 29.7067 & NOMAD\\
                            33011 		& 10 & 315.6318 & 29.7147 & NOMAD\\
                            33012 		& 4  & 315.7199 & 29.8698 & NOMAD\\
                            33021 		& 46 &   0.2883 & 30.0809 & NOMAD\\
                            33022 		& 7  &   0.3767 & 30.2360 & NOMAD\\
                            33023 		& 6  &   0.3994 & 29.9371 & NOMAD\\
                            33031 		& 13 &  44.9427 & 29.7678 & NOMAD\\
                            33032 		& 1  &  45.0308 & 29.9229 & NOMAD\\
                            53001 		& 30 & 324.3935 & 57.1716 & Gaia DR1\\
                            53002 		& 74 & 324.3951 & 57.7616 & Gaia DR1\\
                            53003 		& 13 & 324.4151 & 57.7216 & Gaia DR1\\
                        \enddata
                        \tablenotetext{a}{The names have no special significance, except in some cases they are the name of a recognizable field (trumpler37 and n7789a) and some correspond to the hour angle at which to use them.}
                        \tablenotetext{b}{Not all attempts were successful.}
                        \tablenotetext{c}{The RA/DEC coordinates are for the field center.}
                        \tablenotetext{d}{While the catalog for selecting the targets changed from NOMAD to Gaia DR1, the field identification by PlateMaker continued to use the NOMAD catalog.}
                    \end{deluxetable*}
                
    \section{Results}
    \label{sec:results}
    
        The results in this section are collected from tests of the integrated system required to meet the primary goals of ProtoDESI, including: guiding tests, which required the GFA and telescope, and pointing accuracy and stability tests, which required all subsystems to be working together as described in the object exposure sequence. A summary of these results of these tests, as well as key results from the commissioning phase, are listed in Table \ref{res_table}. 
        
        \subsection{Guiding}
        \label{sec:guiding}
        
            The requirement for the DESI guiding system is that it be capable of delivering a tracking error signal to the TCS better than 100 mas RMS, of which 30 mas come from GFA errors, and of acquiring a guide signal for stars as faint as $r_\mathrm{AB} = 17$. This ensures that at least 10 guide stars will be available in any DESI target FOV, given calculated star densities at any galactic latitude \citep{starcounts}. With a $\mathrm{FOV}\approx 29.3 \, \mathrm{arcmin}^2$, the GFA camera could usually see up to five bright stars ($15 \leq r_\mathrm{AB} \leq 19$).
        
\begin{figure*}
                \centering
                \includegraphics[width=12cm]{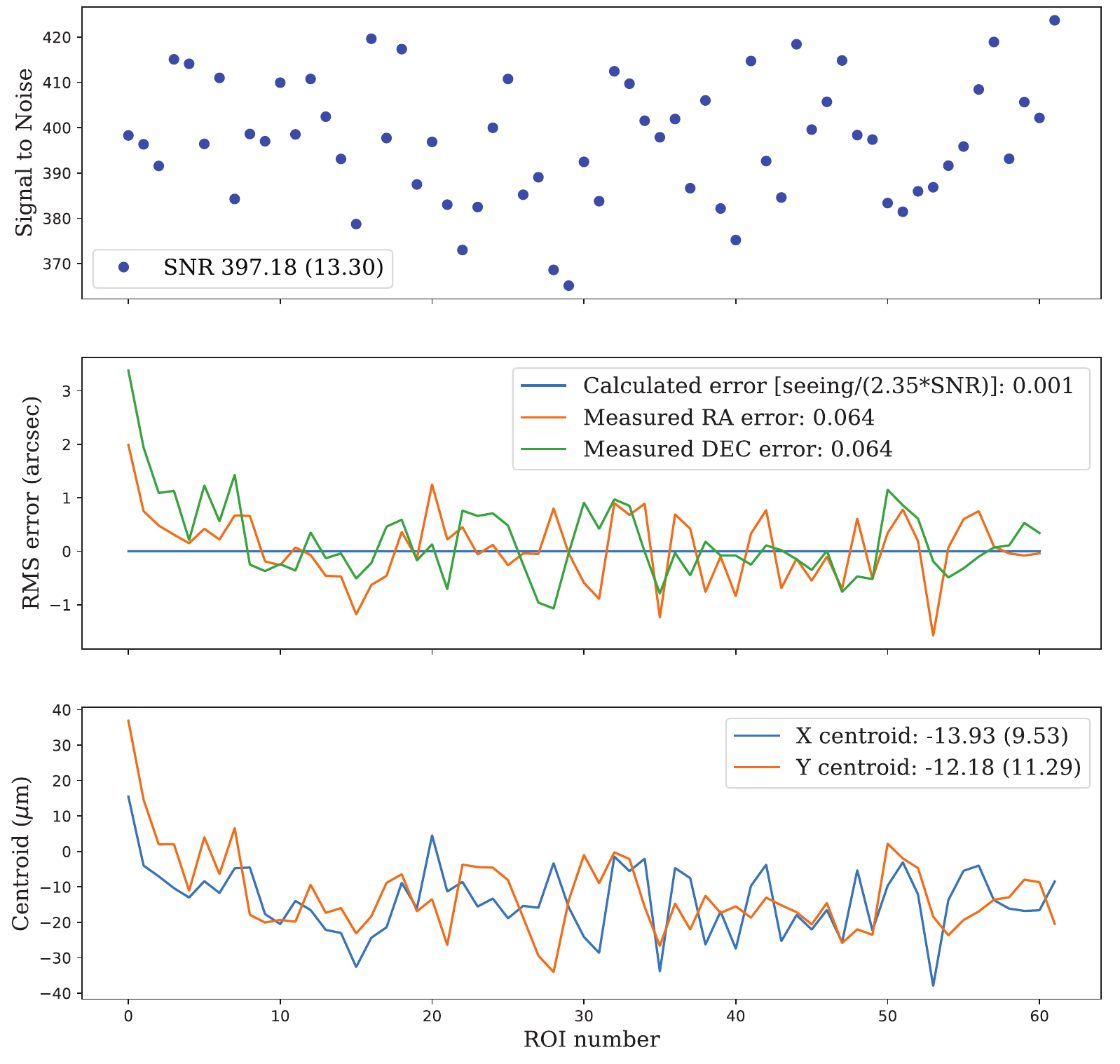}
                \caption{Guiding statistics for a guide star of $r_\mathrm{AB} = 12.27$. All values in parentheses are the RMS errors on the mean. The SNR was sufficiently high (top panel) to yield a statistical centroiding error less than 30 mas RMS (blue line, middle panel), which is the combined RA/DEC error calculated directly from the SNR that would be expected. Despite excellent SNR, the actual correction signals sent to the telescope were much larger than expected (RA/DEC errors in middle panel); the difference is primarily due to atmospheric seeing. The centroid location of the star in the guider images moved quite a bit (bottom panel). There is a clear correlation between the change in centroid location and the measured RMS error, with a plate scale on the focal plate of 0.017 arcsec/$\mu$m.}
                \label{gfa_snr}
            \end{figure*}

            Determining whether or not the ProtoDESI guiding system met the 100 mas requirement was not possible, as the overwhelming majority of the guiding error came from atmospheric seeing. However, based on SNR measurements, the GFA contribution to this error budget easily met the 30 mas requirement, as shown in Fig. \ref{gfa_snr} for a $r_\mathrm{ab} = 12.27$ guide star. The guider correction signals were dominated by motion of the image centroid resulting from atmospheric scintillation, made clear by the correlation between the guiding error and centroid location. Based on these results, DESI will increase the exposure time of the GFA images, and given the large FOV for DESI it is expected that the seeing effects will average out.
            
\begin{figure}
                \centering
                \includegraphics[width=8cm]{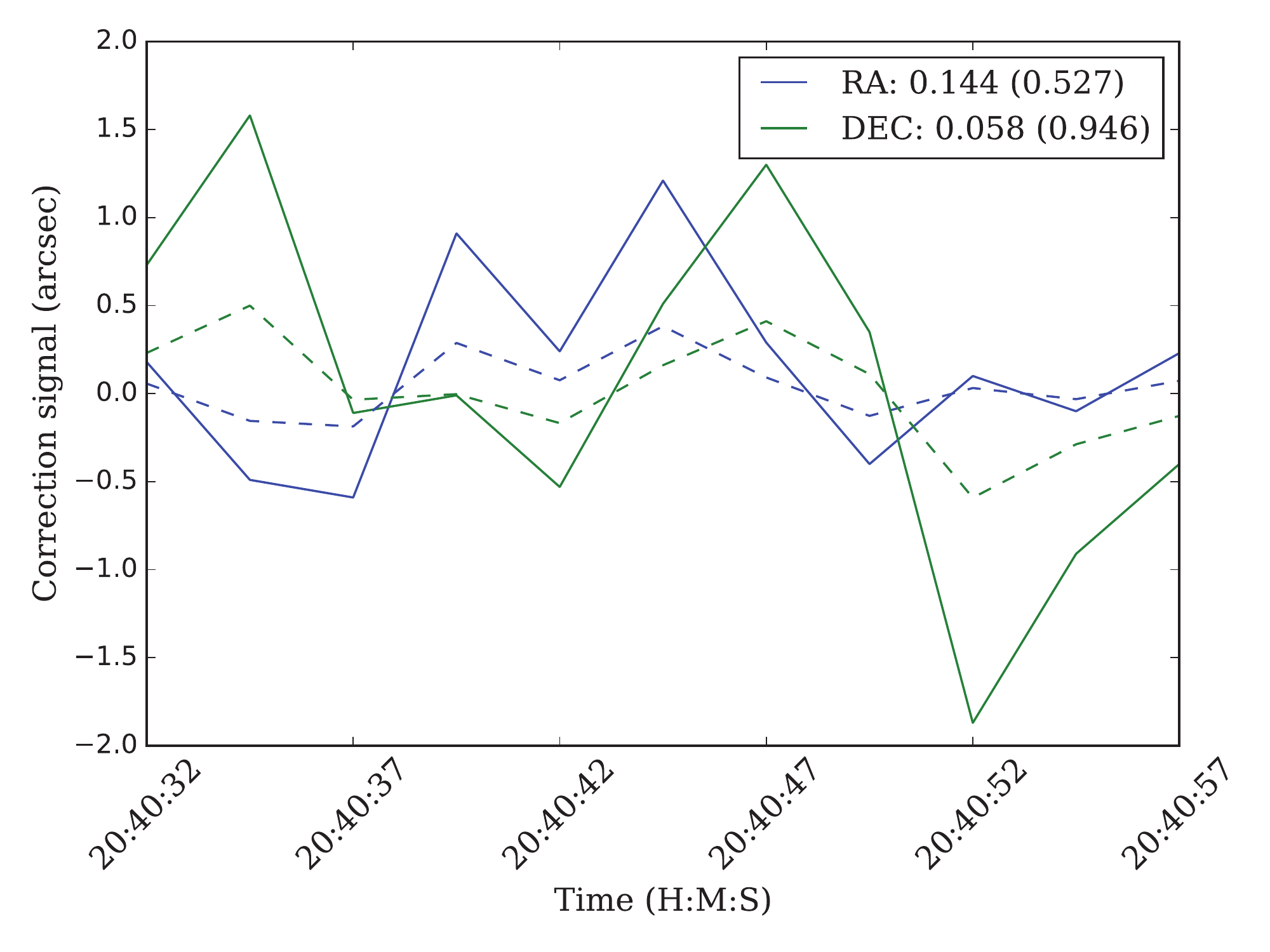}
                \caption{Guiding signal from a 17.09 mag star on Sept. 27, 2016. These guide correction signals were sent to the TCS for $\approx$5 min. on a night with 0.83" seeing. The correction signal was determined from 1-second GFA images, with the mean value listed with the RMS error in parentheses. The solid lines represent the actual measurements, and the dashed lines are the theoretical correction signal from the statistical properties of using 10 guide stars, reducing the RMS error considerably. }
                \label{17mag}
            \end{figure}

\begin{figure*}
                \centering
                \includegraphics[width=17cm]{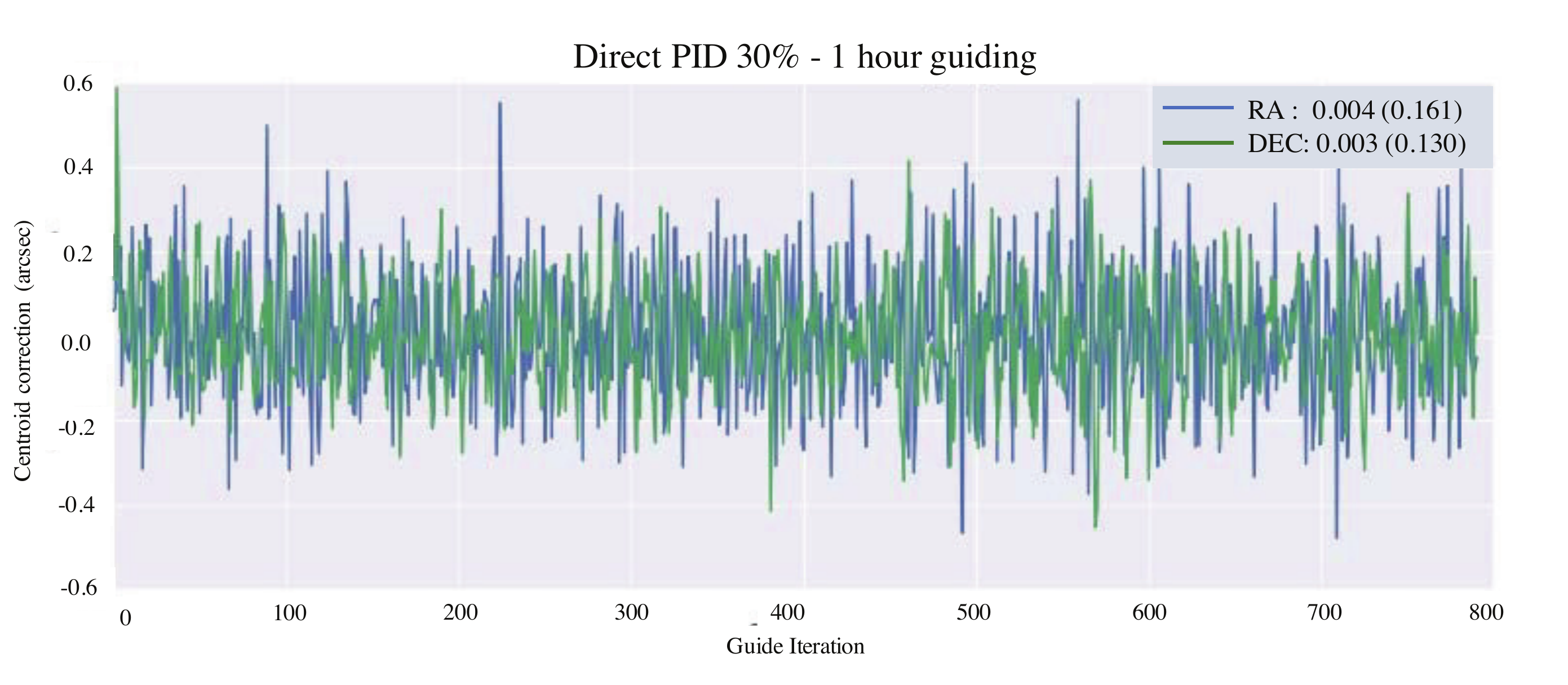}
                \caption{The guiding correction signals sent to the TCS for one hour, about three times the duration of a typical DESI exposure. During this guiding test, the proportional integral derivative (PID) was set to 30\%.}
                \label{long_guide}
            \end{figure*}

            Using the direct guiding mode described above, we tested the GFA guiding sensitivity as a function of star magnitude. The GFA had sufficient SNR on 17 mag stars to acquire a guiding signal (Fig. \ref{17mag}), but failed at magnitudes greater than 17.5. Since DESI will likely have more than 10 guide stars at any time, Fig. \ref{17mag} also shows the expected guiding errors for 10 stars. The guider was run in a direct feedback mode and another mode with a closed loop proportional-integral-derivative (PID). The results of both modes are comparable as we did not have time to tune the PID loop. With additional tuning of the PID loop, changes made to the GFA and guiding software, and increase of exposure time, we are confident that DESI can meet the guiding requirements even with these faint stars. Finally, the guider demonstrated capability of maintaining a guide signal for more than an hour (Fig. \ref{long_guide}).
            
        \subsection{Pointing Accuracy and Stability}
        \label{sec:pointing}
        
\begin{figure*}
                \centering
                \includegraphics[width=12cm]{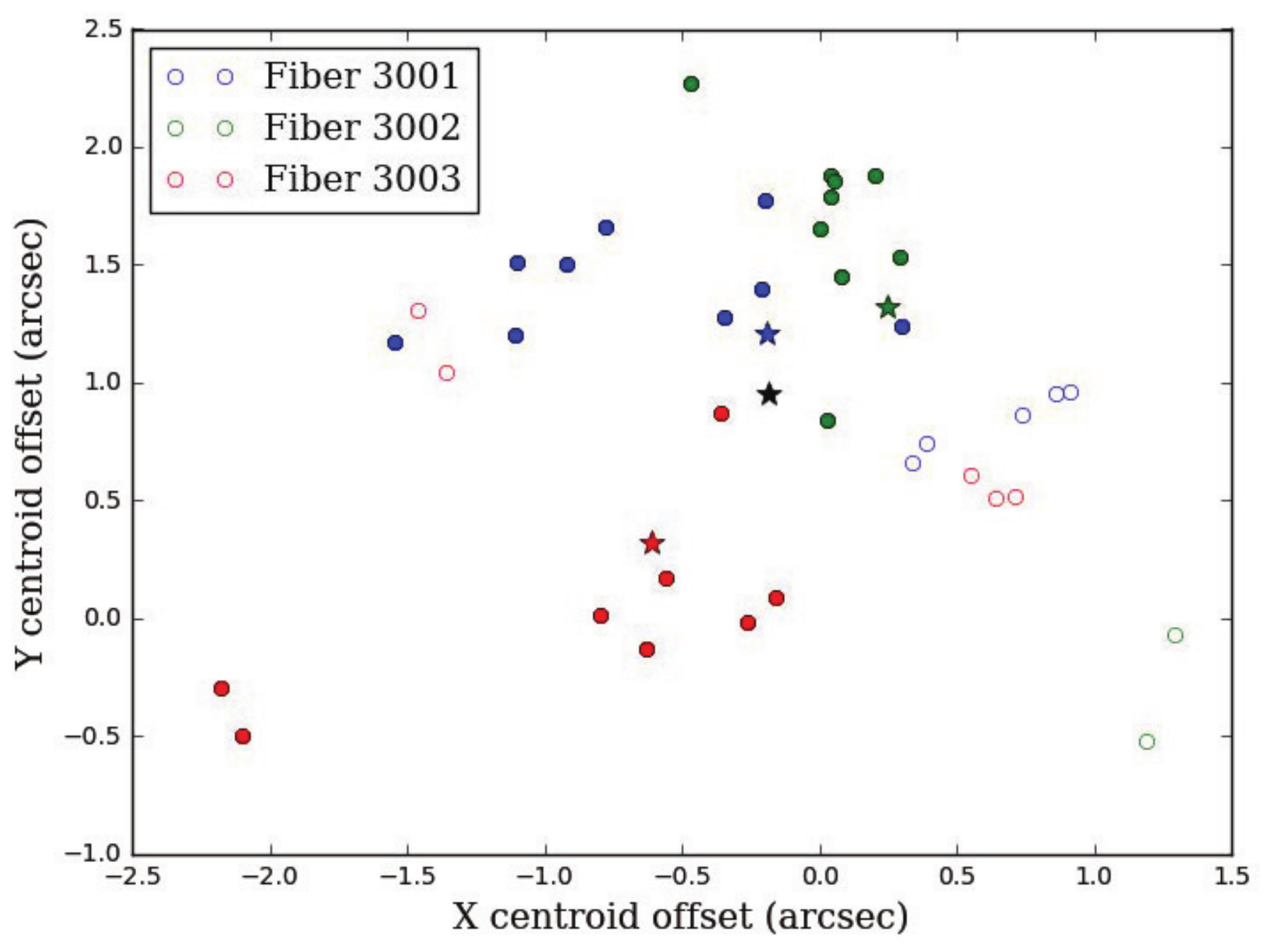}
                \caption{Scatter plot showing the results of successful target acquisition tests to measure positioning accuracy. Each color represents one of the three fibers, and each point is the offset measured from a telescope dither sequence. The filled circles show the results from targets selected from the Gaia DR1 catalog and the open circles are from the NOMAD catalog. The stars are weighted mean offsets for each fiber and the black star is the vector mean offset.}
                \label{pointacc}
            \end{figure*}

            The expectations for pointing were that we align the center of the fiber with the target center within 10 $\mu$m and maintain that pointing accuracy for the duration of a DESI exposure time. While we previously showed that we could point the fibers to within 10 $\mu$m of where they were commanded (Sec. \ref{sec:acctest}), we needed to confirm that this corresponded to actual targets on-sky. To explore how well we were pointing, we dithered the telescope around its initial position as described above. By calculating the intensity-weighted centroid using 25 measurements for each dither grid, we determined where the peak intensity lay in RA/DEC space and the initial pointing offset from the targets. On average, the fibers appeared to be offset from the targets by $1.38\pm 0.30$ arcsec in magnitude for all fibers and successive acquisitions (Fig. \ref{pointacc}).
        
\begin{figure*}[h!]
                \centering
                \includegraphics[width=10cm]{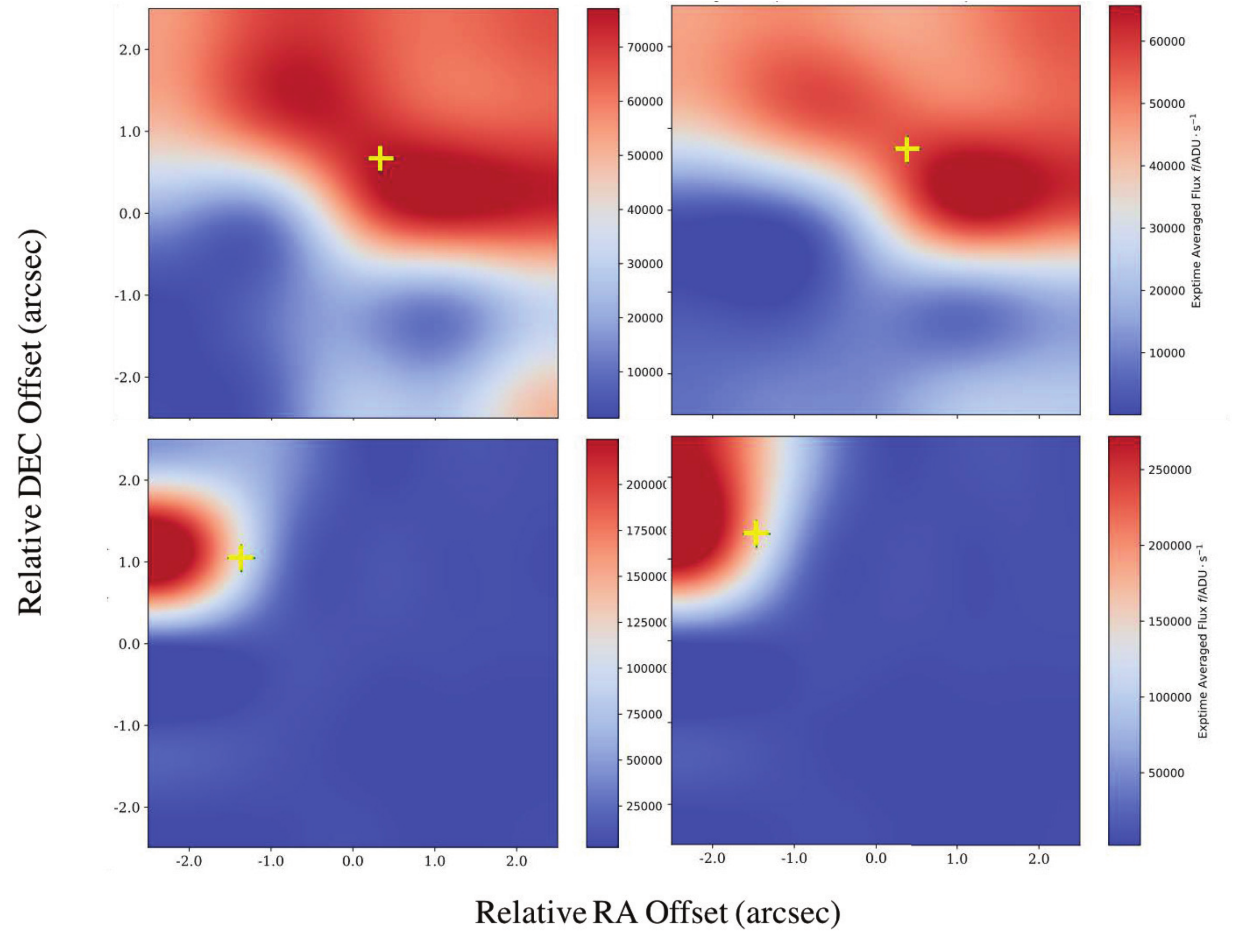}
                \caption{A series of two consecutive $5 \times 5$ telescope dither observations of field trumpler37. The telescope was moved in steps of 1 arcsec and the results for fibers 3001 (top) and 3003 (bottom) are shown. There is a significant difference in focus for the two fibers. There is a systematic offset of 1-2 arcsec, represented by the yellow crosses, from the initial telescope pointing for both positioners. The coordinates for these stars were derived from the NOMAD catalog, whose astrometry may be worse than that of Gaia DR1.}
                \label{onearcsec}
            \end{figure*}

\begin{figure*}[h!]
                \centering
                \includegraphics[width=16cm]{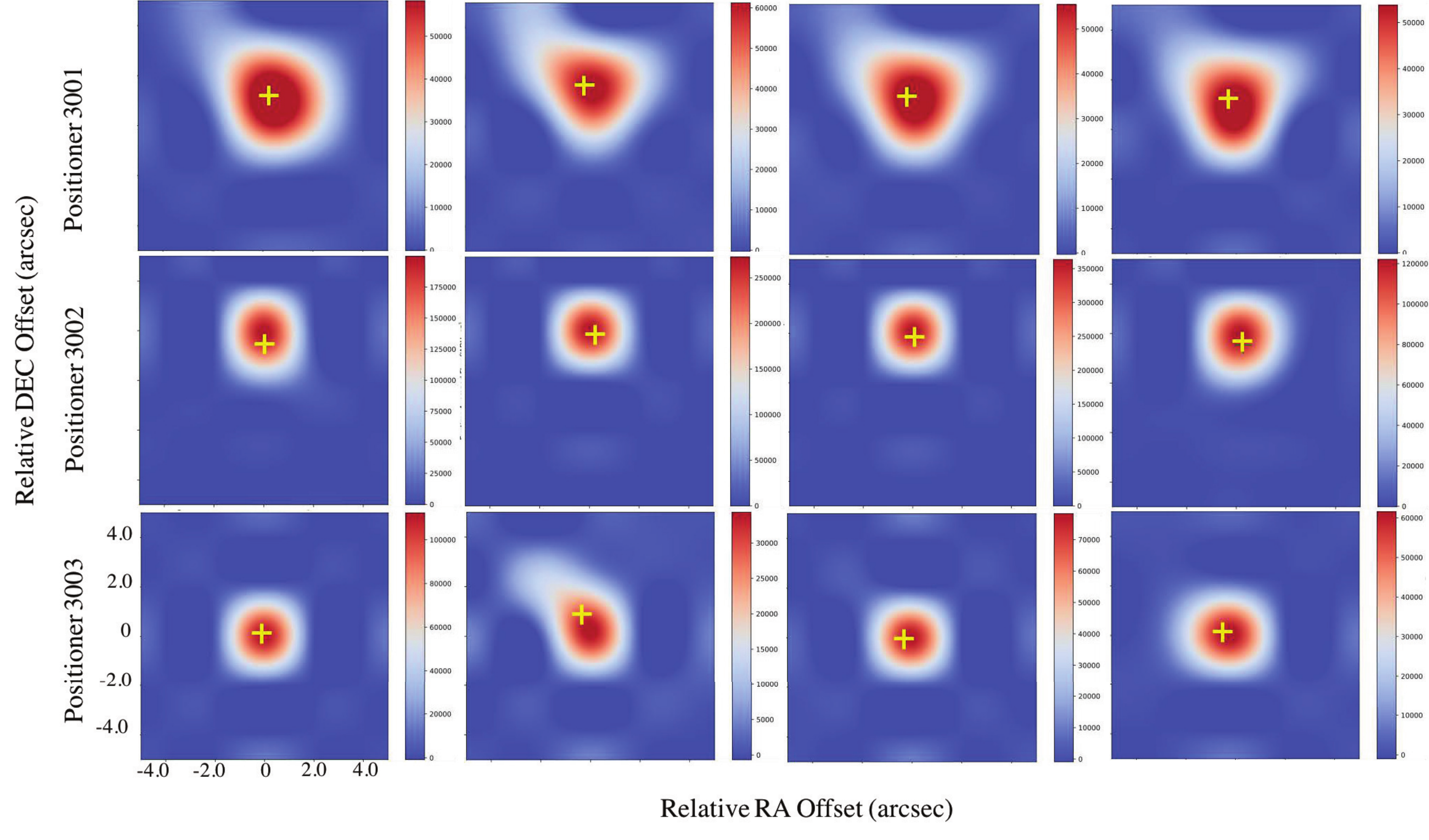}
                \caption{A similar $5 \times 5$ telescope dither on field 53002, with the stellar coordinates derived from the Gaia catalog. Signals from all three fibers are plotted, each column showing the results from a unique object exposure sequence. The acquisition sequence was repeated four times on the same night. The dither steps were 2 arcsec, so we should expect all of the light to be contained in a single tile. The yellow cross is the location of the flux-weighted centroid in each dither sequence, and the color bar show the exposure time-averaged flux in ADU/sec. Fiber 3001 seems out of focus and fiber 3002 is offset by $\approx$2 arcsec. Note that even for a single positioner (row) there were slight differences in the flux from one observation to the next.}
                \label{offset}
            \end{figure*}

            Several factors contributed to these offsets. First, two survey catalogs were used for generating target coordinates, NOMAD and Gaia DR1. It was found that the offset in the positioners decreased slightly when using the astrometry from Gaia DR1, even though the field identification by PlateMaker continued to use the NOMAD catalog. Second, an apparent tilt across the focal plate was identified in GFA images by measuring a differential in the focus across the CCD. To investigate this, telescope dithers were performed at several focus levels and it was found that when the GFA was in focus, which is where we measured the focus, the output of each fiber was at a different focus. This can be seen in Fig. \ref{onearcsec}, which shows two dithering tests results for two fibers, and fiber 3001 (upper panel) is much more spread out than 3003 (lower panel). Fiber 3001 was furthest from the GFA and therefore most affected by the tilt. The target was re-acquired and the dither pattern repeated, showing that this result was reproducible. We returned to certain fields several times in a given night and throughout the weeks to test the repeatability of the offsets. The dithers from these fields gave visibly consistent results for repeat tests (Fig. \ref{offset}), with slight variations in their weighted offset.  This tilt was later confirmed on a CMM to be 0.3 degrees after ProtoDESI was removed from the telescope and shipped back to the laboratory, corresponding to $\Delta z \approx 160 \,\mu\mathrm{m}$ across the GFA and greater than 200 $\mu$m focus error for the positioner placed furthest from the GFA. 
            
\begin{figure*}
                \centering
                \includegraphics[width=12cm]{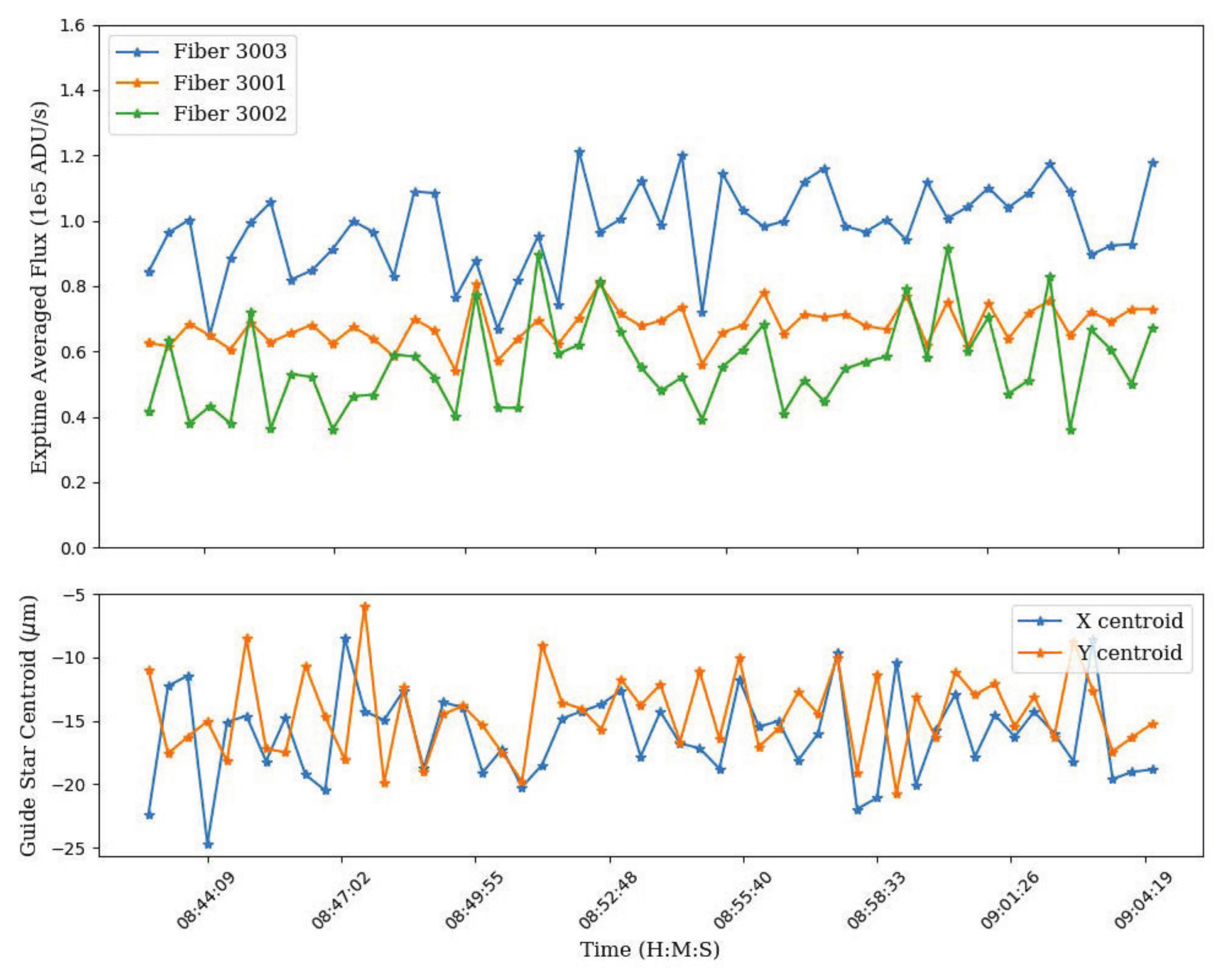}
                \caption{Stability tests for all three fibers on September 25, 2016 for field 53002. Each fiber was aligned with a different target and 10-second FPC images were taken in succession over a 30-minute period. The measured flux fluctuates up to 25\% over this time period. We again see correlations between the averaged GFA guide star centroid location and the fluctuations in the target flux.}
                \label{stability}
            \end{figure*}

            The pointing stability of a ProtoDESI observation depended on both the pointing accuracy and stability of a fiber positioner and the guiding accuracy of the combined GFA+telescope system. In order to assess the pointing stability, we first acquired a target and then took a series of 10 second FPC images, two per minute, for a period of 20 minutes (corresponding to a typical DESI exposure duration). The fibers remained at the initial target acquisition with no offsets applied. The results showed fluctuations of $\approx$30\% in fiber spot intensity around the mean value, with some 10-second integrations resulting in fiber spots 70\% below the peak value. This may be the result of imperfect initial target acquisition, i.e., where the fiber was not centered on the target star, but instead on the wings of its PSF. As a consequence, the normal atmospheric seeing fluctuations resulted in a larger-than-expected spot intensity variation. In fact, we saw that the guide star centroids vary considerably over this time period after averaging over the 10 second FPC exposure times (Fig. \ref{stability}).
        
            \begin{deluxetable*}{cccc}
                        \tablecaption{Key ProtoDESI Requirements and Results \label{res_table}}          
                        \tablecolumns{4}
                        \tablehead{
                        \colhead{Description} &
                        \colhead{Requirement} & 
                        \colhead{Result} &
                        \colhead{Ref.}
                        }
                        \startdata
                            Positioner Accuracy\tablenotemark{a} & $\leq$10 $\mu$m & 6 $\mu$m &  \ref{sec:acctest}\\
                            FVC Centroid Precision\tablenotemark{b} & 3 $\mu$m & $\leq$3 $\mu$m & \ref{sec:fvc_centroid}\\
                            FPC Relative Photometry & 1$\%$ & $\leq$0.8\% &  \ref{sec:fpc_results}\\
                            GFA Centroid Accuracy\tablenotemark{c} & 30 mas (@ 1Hz) & $\approx$10 mas & \ref{sec:guiding} \\
                            Guiding Sensitivity ($r_{AB}$) & $\leq$17 & 17.5 & \ref{sec:guiding}\\
                            Pointing Accuracy\tablenotemark{d} & 0.17 arcsec & 1.38 $\pm$ 0.3 arcsec & \ref{sec:pointing}\\
                        \enddata
                        \tablenotetext{a}{Test on-telescope with dome open.}
                        \tablenotetext{b}{Requirement met at exposure times as low as 0.5 sec.}
                        \tablenotetext{c}{Requirement changed to a lower frequency.}
                        \tablenotetext{d}{See Sec. \ref{sec:disc} for discussion of the discrepancy between requirement and result.}
                    \end{deluxetable*}


    \section{Discussion}
    \label{sec:disc}
    
        The scope and scale of the investment in the DESI project necessitate a plan to retire risks associated with achieving its science goals. This is achievable during the development stage by targeting hardware and software interfaces lying on the critical path, representing both technical and programmatic risks. ProtoDESI's construction and operations were conducted with a small team, due in large part to the desire to impact the overall DESI project and schedule as minimally as possible. This resulted in two issues: some of the hardware was not in its final performance condition; and the team was over-stretched in terms of personnel and working hours. Despite these challenges, ProtoDESI exemplifies an effective approach to exercising interfaces, gaining operational experience, and identifying improvements to the instrument on a useful time scale.  Furthermore, a small dedicated team was able to uncover and resolve problems, as well as retiring risks associated with system performance, and build a basis for commissioning plans.
    
        This experiment was designed to test the pointing accuracy and stability of the robotic fiber positioners and the target acquisition and guiding capabilities of the GFAs. We successfully demonstrated these to varying levels of accuracy and were able to recommend changes in design or operations for the larger DESI project. Key improvements will be made to the GFA and its operation, including the GFA thermal system, electronics design issues associated with the ethernet link, and requiring the GFA to send full frame images to the guider software at a rate of 0.1 Hz rather than ROIs at 1 Hz. Other hardware design improvements will include mechanical bolstering of the FVC lens and modifications to the robotic positioners. Operational and process oriented issues were also identified during ProtoDESI and have already impacted the development of DESI assembly, integration, and test procedures and ongoing commissioning planning. These include early development of data models and interfaces, clear plans for precision metrology integrated across subsystems, and lab-testing of the instrument in every configuration expected to be used on the telescope. 
    
        There are several sources that contributed to our pointing error: focal plate tilt, FPC DC offset variations, atmospheric turbulence, astrometric errors, and metrology errors. While each of these will be addressed individually, their combined impact will be apparent only after DESI is installed. However, the accuracy achieved with ProtoDESI is well within expectations given the limitations on metrology and incomplete characterization of the Mosaic corrector and FVC optical distortions. The DESI team is confident that with improvements, DESI will achieve its science goals.

    \section{Conclusion}
    \label{sec:conc}
    
        DESI, a fiber-fed spectrograph to measure the spectra of up to 35 million galaxies and quasars, will be integrated on the Mayall telescope in 2018. Featuring 5,000 robotic positioners to align optical fibers with targets simultaneously, DESI will be vastly more efficient than its predecessors in data collection. The ProtoDESI experiment was built and commissioned at the Mayall telescope in the summer of 2016 to test the capability of the DESI hardware and software to acquire targets and maintain stability, serving as the first on-sky demonstration of critical DESI technology. ProtoDESI demonstrated the ability to complete an object exposure sequence, which required guiding the telescope, maintaining stability, and acquiring targets with the robotic fiber positioners. Guiding was successful on stars as faint as $r_{\mathrm{AB}} = 17$ mag, and the final guiding errors were consistent with atmospheric seeing ($>$ 100 mas). On average, the final position of the positioners were 1.38 $\pm$ 0.30 arcsec from the center of the target, and this accuracy was within expectations given the ProtoDESI setup. The success of ProtoDESI provided key project members with experience that will facilitate a smooth and efficient commissioning period for DESI, and gave us confidence that DESI will achieve its science goals.
    
        \acknowledgments
        
            This research is supported by the Director, Office of Science, Office of High Energy Physics of the U.S. Department of Energy under Contract No. DE–AC02–05CH1123, and by the National Energy Research Scientific Computing Center, a DOE Office of Science User Facility under the same contract; additional support for DESI is provided by the U.S. National Science Foundation, Division of Astronomical Sciences under Contract No. AST-0950945 to the National Optical Astronomy Observatory; the Science and Technologies Facilities Council of the United Kingdom; the Gordon and Betty Moore Foundation; the Heising-Simons Foundation; the National Council of Science and Technology of Mexico, and by the DESI Member Institutions.
        
            The research leading to these results has also received funding from the European Research Council under the European Union’s Seventh Framework Programme (FP/2007- 2013) / ERC Grant Agreement n. 320964 (WDTracer).
        
            This work has made use of data from the European Space Agency (ESA) mission {\it Gaia} (\url{https://www.cosmos.esa.int/gaia}), processed by the {\it Gaia} Data Processing and Analysis Consortium (DPAC, \url{https://www.cosmos.esa.int/web/gaia/dpac/consortium}). Funding for the DPAC has been provided by national institutions, in particular the institutions participating in the {\it Gaia} Multilateral Agreement.
        
            This work was based on observations at Kitt Peak National Observatory, National Optical Astronomy Observatory, which is operated by the Association of Universities for Research in Astronomy (AURA) under cooperative agreement with the National Science Foundation. The authors are honored to be permitted to conduct astronomical research on Iolkam Du’ag (Kitt Peak), a mountain with particular significance to the Tohono O'odham Nation.
    \newpage 
    \bibliographystyle{aasjournal}
    \bibliography{pd_paper}

\begin{thebibliography}{}
\expandafter\ifx\csname natexlab\endcsname\relax\def\natexlab#1{#1}\fi
\providecommand{\url}[1]{\href{#1}{#1}}

\bibitem[{Abareshi {et~al.}(2016)Abareshi, Marshall, Gott, Sprayberry,
  Cantarutti, Joyce, Williams, Probst, Reetz, Paat, Butler, Soto, Dey, \&
  Summers}]{mayall1}
Abareshi, B., Marshall, R., Gott, S., {et~al.} 2016, Proc. SPIE, 9913,
  doi:10.1117/12.2233087

\bibitem[{{Albrecht} {et~al.}(2006){Albrecht}, {Bernstein}, {Cahn}, {Freedman},
  {Hewitt}, {Hu}, {Huth}, {Kamionkowski}, {Kolb}, {Knox}, {Mather}, {Staggs},
  \& {Suntzeff}}]{DETF}
{Albrecht}, A., {Bernstein}, G., {Cahn}, R., {et~al.} 2006, ArXiv Astrophysics
  e-prints, astro-ph/0609591

\bibitem[{Bahcall(1986)}]{starcounts}
Bahcall, J.~N. 1986, Annual Review of Astronomy and Astrophysics, 24, 577

\bibitem[{{Dawson} {et~al.}(2013){Dawson}, {Schlegel}, {Ahn}, {Anderson},
  {Aubourg}, {Bailey}, {Barkhouser}, {Bautista}, {Beifiori}, {Berlind},
  {Bhardwaj}, {Bizyaev}, {Blake}, {Blanton}, {Blomqvist}, {Bolton}, {Borde},
  {Bovy}, {Brandt}, {Brewington}, {Brinkmann}, {Brown}, {Brownstein}, {Bundy},
  {Busca}, {Carithers}, {Carnero}, {Carr}, {Chen}, {Comparat}, {Connolly},
  {Cope}, {Croft}, {Cuesta}, {da Costa}, {Davenport}, {Delubac}, {de Putter},
  {Dhital}, {Ealet}, {Ebelke}, {Eisenstein}, {Escoffier}, {Fan}, {Filiz Ak},
  {Finley}, {Font-Ribera}, {G{\'e}nova-Santos}, {Gunn}, {Guo}, {Haggard},
  {Hall}, {Hamilton}, {Harris}, {Harris}, {Ho}, {Hogg}, {Holder}, {Honscheid},
  {Huehnerhoff}, {Jordan}, {Jordan}, {Kauffmann}, {Kazin}, {Kirkby}, {Klaene},
  {Kneib}, {Le Goff}, {Lee}, {Long}, {Loomis}, {Lundgren}, {Lupton}, {Maia},
  {Makler}, {Malanushenko}, {Malanushenko}, {Mandelbaum}, {Manera}, {Maraston},
  {Margala}, {Masters}, {McBride}, {McDonald}, {McGreer}, {McMahon}, {Mena},
  {Miralda-Escud{\'e}}, {Montero-Dorta}, {Montesano}, {Muna}, {Myers},
  {Naugle}, {Nichol}, {Noterdaeme}, {Nuza}, {Olmstead}, {Oravetz}, {Oravetz},
  {Owen}, {Padmanabhan}, {Palanque-Delabrouille}, {Pan}, {Parejko},
  {P{\^a}ris}, {Percival}, {P{\'e}rez-Fournon}, {P{\'e}rez-R{\`a}fols},
  {Petitjean}, {Pfaffenberger}, {Pforr}, {Pieri}, {Prada}, {Price-Whelan},
  {Raddick}, {Rebolo}, {Rich}, {Richards}, {Rockosi}, {Roe}, {Ross}, {Ross},
  {Rossi}, {Rubi{\~n}o-Martin}, {Samushia}, {S{\'a}nchez}, {Sayres}, {Schmidt},
  {Schneider}, {Sc{\'o}ccola}, {Seo}, {Shelden}, {Sheldon}, {Shen}, {Shu},
  {Slosar}, {Smee}, {Snedden}, {Stauffer}, {Steele}, {Strauss}, {Streblyanska},
  {Suzuki}, {Swanson}, {Tal}, {Tanaka}, {Thomas}, {Tinker}, {Tojeiro},
  {Tremonti}, {Vargas Maga{\~n}a}, {Verde}, {Viel}, {Wake}, {Watson}, {Weaver},
  {Weinberg}, {Weiner}, {West}, {White}, {Wood-Vasey}, {Yeche}, {Zehavi},
  {Zhao}, \& {Zheng}}]{BOSS}
{Dawson}, K.~S., {Schlegel}, D.~J., {Ahn}, C.~P., {et~al.} 2013, \aj, 145, 10

\bibitem[{{DESI Collaboration} {et~al.}(2016{\natexlab{a}}){DESI
  Collaboration}, {Aghamousa}, {Aguilar}, {Ahlen}, {Alam}, {Allen}, {Allende
  Prieto}, {Annis}, {Bailey}, {Balland}, \& et~al.}]{SDR}
{DESI Collaboration}, {Aghamousa}, A., {Aguilar}, J., {et~al.}
  2016{\natexlab{a}}, ArXiv e-prints, arXiv:1611.00036

\bibitem[{{DESI Collaboration} {et~al.}(2016{\natexlab{b}}){DESI
  Collaboration}, {Aghamousa}, {Aguilar}, {Ahlen}, {Alam}, {Allen}, {Allende
  Prieto}, {Annis}, {Bailey}, {Balland}, \& et~al.}]{IDR}
---. 2016{\natexlab{b}}, arXiv:1611.00037v2

\bibitem[{Dey {et~al.}(2016)Dey, Rabinowitz, Karcher, Bebek, Baltay,
  Sprayberry, Valdes, Stupak, Donaldson, Emmet, Hurteau, Abareshi, Marshall,
  Lang, Fitzpatrick, Daly, Joyce, Schlegel, Schweiker, Allen, Blum, \&
  Levi}]{Mosaic3}
Dey, A., Rabinowitz, D., Karcher, A., {et~al.} 2016, Proc. SPIE 9908

\bibitem[{{Gaia Collaboration} {et~al.}(2016){Gaia Collaboration}, {Brown},
  {Vallenari}, {Prusti}, {de Bruijne}, {Mignard}, {Drimmel}, {Babusiaux},
  {Bailer-Jones}, {Bastian}, \& et~al.}]{Gaia}
{Gaia Collaboration}, {Brown}, A.~G.~A., {Vallenari}, A., {et~al.} 2016, \aap,
  595, A2

\bibitem[{Honscheid {et~al.}(2010)Honscheid, Eiting, Elliott, Annis, Bonati,
  Buckley-Geer, Castander, da~Costa, Haney, Hanlon, Karliner, Kuehn, Kuhlmann,
  Marshall, Meyer, Neilsen, Ogando, Roodman, Schalk, Schumacher, Selen,
  Serrano, Thaler, \& Wester}]{DES}
Honscheid, K., Eiting, J., Elliott, A., {et~al.} 2010, Proc. SPIE, 7740,
  doi:10.1117/12.856734

\bibitem[{Honscheid {et~al.}(2016)Honscheid, Elliott, Beaufore, Buckley-Geer,
  Castander, daCosta, Fausti, Kent, Kirkby, Neilsen, Reil, Serrano, \&
  Slozar}]{spie_ics}
Honscheid, K., Elliott, A.~E., Beaufore, L., {et~al.} 2016, Proc. SPIE, 9913,
  doi:10.1117/12.2229835

\bibitem[{Jacoby {et~al.}(1998)Jacoby, Liang, Vaughnn, Reed, \&
  Armandroff}]{doi:10.1117/12.316845}
Jacoby, G.~H., Liang, M., Vaughnn, D., Reed, R., \& Armandroff, T. 1998, Proc.
  SPIE, 3355, doi:10.1117/12.316845

\bibitem[{{Kaiser}(1987)}]{kaiser}
{Kaiser}, N. 1987, \mnras, 227, 1

\bibitem[{Kent {et~al.}(2016)Kent, Lampton, Doel, Brooks, Miller, Besuner,
  Silber, Liang, Sprayberry, Baltay, \& Rabinowitz}]{spie_platemaker}
Kent, S., Lampton, M., Doel, A.~P., {et~al.} 2016, Proc. SPIE, 9908,
  doi:10.1117/12.2232689

\bibitem[{{Kent}(2017)}]{kentprep}
{Kent}, S.~M. 2017, ArXiv e-prints, arXiv:1711.03916

\bibitem[{{Kleinmann} {et~al.}(1994){Kleinmann}, {Lysaght}, {Pughe},
  {Schneider}, \& {Skrutskie}}]{2mass}
{Kleinmann}, S.~G., {Lysaght}, M.~G., {Pughe}, W.~L., {Schneider}, S.~E., \&
  {Skrutskie}, M.~F. 1994, NASA STI/Recon Technical Report N, 95

\bibitem[{{Kuehn} {et~al.}(2013){Kuehn}, {Kuhlmann}, {Allam}, {Annis},
  {Bailey}, {Balbinot}, {Bernstein}, {Biesiadzinski}, {Burke}, {Butner},
  {Camargo}, {da Costa}, {DePoy}, {Diehl}, {Dietrich}, {Estrada}, {Fausti},
  {Gerke}, {Guarino}, {Head}, {Kessler}, {Lin}, {Lorenzon}, {Maia}, {Maki},
  {Marshall}, {Nord}, {Neilsen}, {Ogando}, {Park}, {Peoples}, {Rastawicki},
  {Rheault}, {Santiago}, {Schubnell}, {Seitzer}, {Smith}, {Spinka},
  {Sypniewski}, {Tarle}, {Tucker}, {Walker}, \& {Wester}}]{precam}
{Kuehn}, K., {Kuhlmann}, S., {Allam}, S., {et~al.} 2013, \pasp, 125, 409

\bibitem[{Schubnell {et~al.}(2016)Schubnell, Ameel, Besuner, Gershkovich,
  Heetderks, Hoerler, Kneib, Heetderks, Silber, Tarlé, \&
  Weaverdyck}]{spie_positioners}
Schubnell, M., Ameel, J., Besuner, R.~W., {et~al.} 2016, Proc. SPIE, 9908,
  doi:10.1117/12.2233370

\bibitem[{{Sebag} {et~al.}(2014){Sebag}, {Gressler}, {Neill}, {Barr}, {Claver},
  \& {Andrew}}]{lsst_comcam}
{Sebag}, J., {Gressler}, W., {Neill}, D., {et~al.} 2014, in \procspie, Vol.
  9145, Ground-based and Airborne Telescopes V, 91454A

\bibitem[{Seo \& Eisenstein(2003)}]{seo}
Seo, H.-J., \& Eisenstein, D.~J. 2003, The Astrophysical Journal, 598, 720

\bibitem[{Sprayberry {et~al.}(2016)Sprayberry, Dunlop, Evatt, Reddell, Gott,
  George, Donaldson, Stupak, Marshall, Abareshi, Stover, Warner, Cantarutti, \&
  Probst}]{mayall2}
Sprayberry, D., Dunlop, P., Evatt, M., {et~al.} 2016, Proc. SPIE, 9906,
  doi:10.1117/12.2233177

\bibitem[{{Tuttle} {et~al.}(2016){Tuttle}, {Hill}, {Vattiat}, {Lee}, {Drory},
  {Kelz}, {Ramsey}, {Peterson}, {Noyola}, {DePoy}, {Marshall}, {Chonis},
  {Dalton}, {Fabricius}, {Farrow}, {Good}, {Haynes}, {Indahl}, {Jahn}, {Kriel},
  {Nicklas}, {Montesano}, {Prochaska}, {Allen}, {Landriau}, {MacQueen}, {Roth},
  {Savage}, \& {Snigula}}]{hetdex}
{Tuttle}, S.~E., {Hill}, G.~J., {Vattiat}, B.~L., {et~al.} 2016, in \procspie,
  Vol. 9908, Ground-based and Airborne Instrumentation for Astronomy VI, 99081I

\bibitem[{{Weinberg} {et~al.}(2013){Weinberg}, {Mortonson}, {Eisenstein},
  {Hirata}, {Riess}, \& {Rozo}}]{weinberg}
{Weinberg}, D.~H., {Mortonson}, M.~J., {Eisenstein}, D.~J., {et~al.} 2013,
  \physrep, 530, 87

\end{thebibliography}

    \clearpage
    \appendix
    
    \section{Acronyms \& Abbreviations}
    
        \begin{tabular}{ll}
            ADU & Analog-Digital Units\\
            BAO & Baryon Acoustic Oscillations\\
            CCD & Charge Coupled Device\\
            CMM & Coordinate Measuring Machine\\
            DEC & Declination\\
            DES & Dark Energy Survey\\
            DESI & Dark Energy Spectroscopic Instrument\\
            DETF & Dark Energy Task Force\\
            DOS & DESI Online System\\
            DTS & Data Transfer System\\
            FIF & Field Illuminated Fiducial\\
            FOV & Field of View\\
            FPC & Fiber Photometry Camera\\
            FPGA & Field-Programmable Field Array\\
            FVC & Fiber View Camera\\
            FWHM & Full Width Half Maximum\\
            GFA & Guide, Focus and Alignment camera\\
            GIF & GFA Illuminated Fiducial\\
            ICS & Instrument Control System\\
            KPNO & Kitt Peak National Observatory\\
            LBNL & Lawrence Berkeley National Laboratory\\
            MzLS & Mayall z-band Legacy Survey\\
            NOAO & National Optical Astronomy Observatory\\
            NOMAD & Naval Observatory Merged Astrometric Dataset\\
            OCS & Observation Control System\\
            PC  & Petal Controller\\
            PID & Proportional Integral Derivative\\
            PSF & Point Spread Function\\
            QE & Quantum Efficiency\\
            RA  & Right Ascension\\
            RMS & Root Mean Squared\\
            ROI & Region of Interest\\
            RSD & Redshift Space Distortions\\
            SNR & Signal to Noise Ratio\\
            TCS & Telescope Control System\\
        \end{tabular}

\end{document}